\documentclass[referee]{aa}
\usepackage{natbib}
\usepackage{amsmath}
\usepackage{amssymb}
\usepackage{graphicx}
\usepackage[varg]{txfonts}

\begin{document}

\title{The Missing Magnetic Morphology Term in Stellar Rotation Evolution}
\author{ Cecilia Garraffo,  Jeremy J. Drake,  Ofer Cohen}


\institute{Harvard-Smithsonian Center for Astrophysics, 60 Garden St. Cambridge, MA 02138 }


\abstract{}
{This study examines  the relationship between magnetic field complexity and mass and angular momentum losses.  Observations of open clusters have revealed a bimodal distribution of the rotation periods of solar-like stars that has proven difficult to explain under the existing rubric of magnetic braking.
Recent studies suggest that magnetic complexity can play an important role in controlling stellar spin-down rates. However, magnetic morphology is still neglected in most rotation evolution models due to the difficulty of properly accounting for its effects on wind driving and angular momentum loss.}   {Using state-of-the-art magnetohydrodynamical magnetized wind simulations we study the effect that different distributions of the magnetic flux at different levels of geometrical complexity have on mass and angular momentum loss rates. }{Angular momentum loss rates depend strongly on the level of complexity of the field but are independent of the way this complexity is distributed.  We deduce the analytical terms representing the magnetic field morphology dependence of mass and angular momentum loss rates.  We also define a parameter that best represents complexity for real stars. As a test, we use these analytical methods to estimate mass and angular momentum loss rates for $8$ stars with observed magnetograms and compare them to the simulated results.}{Magnetic field complexity provides a natural physical basis for stellar rotation evolution models requiring a rapid transition between weak and strong spin-down modes.}



\authorrunning{Garraffo et al.}
\maketitle

\section{INTRODUCTION}
\label{sec:Intro}

One of the fundamental observable characteristics of a star is its rotation period.  Stellar rotation evolves over time as a result of interior structural adjustments as stars settle onto the main sequence, and as a result of angular momentum loss through magnetized stellar winds \citep[e.g.][]{Schatzman:62,Kraft:67,Weber.Davis:67,Mestel:68,Endal.Sofia:81,Kawaler:88}.  However, our present understanding of the details of this behavior is far from complete.   
Observations of young open clusters  \cite[e.g.][see \citealt{Meibom.etal:11} for a recent compilation]{Stauffer.Hartmann:87, Soderblom.etal:93, Queloz.etal:98, Terndrup.etal:00} have revealed a bimodal distribution of fast and slower rotation rates that has proven difficult to explain with current spin-down models  \cite[e.g.][]{Stauffer.etal:84,Soderblom.etal:93, Barnes:03}.  The situation has been summarized recently by \citet{Brown:14}.

Recently, the morphology of the magnetic field (by morphology we mean the distribution of the magnetic field on the stellar surface, which some authors call topology) has received a lot of attention in the context of stellar rotation evolution \citep{Brown:14, Reville.etal:15a, Garraffo.etal:15, Reville.etal:15b}.  Most previous spin down models had assumed dipolar morphology, with few exceptions that explored the role of the multipole order of the magnetic field \citep{Weber.Davis:67, Mestel.Paris:84, Kawaler:88} but were limited to the effect of the radial dependence of the magnetic field.  A growing number of studies based on ZDI maps indicate that young, active stars have a larger fraction of their magnetic energy stored in higher order multipole components  \citep[e.g.][]{Donati:03, Donati.Landstreet:09, Marsden.etal:11a, Waite.etal:11, Waite.etal:15, Folsom.etal:16}.    \cite{Linsky.Wood:14} have also recently inferred mass loss rates for the active stars $\xi$~Boo~A and $\pi^1$~UMa that are two orders of magnitude lower than expected based on extrapolation from lower activity stars, suggesting that magnetic topology could have a more profound effect on angular momentum loss than simply through the radial field strength dependence. 

\citet[][from here on CG15]{Garraffo.etal:15} used detailed three-dimensional magnetohydrodynamic (MHD) stellar wind simulations to show that a factor representing magnetic morphology, missing in rotation evolution models, can have a drastic effect on mass and angular momentum losses.  They drew a connection between magnetic morphology and the ``metastable dynamo" proposal of \citet{Brown:14} in which bimodal rotation distributions at early ages are attributed to an initially weak coupling of magnetic field to the stellar wind that ``spontaneously and randomly'' changes to a strongly coupled mode and initiation of rapid spin-down.  CG15 suggested that magnetic morphology could provide such a mode switch.  That work was based on a limited set of cases of magnetic complexity assuming idealized magnetic field distributions.  

Here, we investigate the effects of different flux distributions (different $m$) of each term $n$ in a spherical harmonic multipolar expansion representation of the surface magnetic field. Based on these complete study of magnetic flux complexity and distribution, we derive the analytical dependence of mass and angular momentum loss rates on magnetic complexity, which provides the means to realistically estimate those rates without the need of computationally expensive simulations.   

The numerical methods are described in Section~\ref{sec:Methods} and the results of model calculations in 
Section~\ref{sec:Results}. We state our main findings in Section~\ref{sec:Conclusions} .


\section{NUMERICAL SIMULATION}
\label{sec:Methods}

\subsection{MHD model} 
\label{sec:Model}

Our computational magnetohydrodynamic (MHD) procedure follows closely that of \cite{Garraffo.etal:15}; we describe it here only in brief.
We model the stellar wind and coronal structure using the BATS-R-US code \citep{Powell:99,Toth.etal:12}. This code solves the set of non-ideal MHD
Equations --- the coupled conservation laws for mass, momentum, magnetic flux and energy. We use a stretched spherical grid which is dynamically refined to resolve evolved current sheets in the simulation domain. We drive the model with synoptic maps of the photospheric radial magnetic field (magnetograms) that serve as the inner boundary condition for the magnetic field, where this boundary condition is extrapolated analytically using the potential field method, assuming that the field is purely radial at a distance of $r = 4.5R_{\*}$ (the
''source surface'';  \citealt{Altschuler.Newkirk:69}. This procedure enables us to obtain the initial condition for the three-dimensional magnetic field in the volume. Once the initial conditions are set, the model provides self-consistently the additional coronal heating, stellar wind acceleration, and field line forcing by the expanding wind as it accounts for many physics-based processes, such as Alfv\'en wave turbulence, radiative cooling, and electron heat conduction, all in a self-consistent manner (see
\citealt{Oran.etal:13, Sokolov.etal:13, Vanderholst.etal:14}, for full details). 

Unlike models
that were used to study stellar winds in which the solutions were brought to a pressure balance between the prescribed, spherical "Parker" wind and the potential magnetic field (e.g., \citealt{Matt.etal:12,Vidotto.etal:14b}), here we account for the additional acceleration of the wind and the role of the magnetic field in accelerating the wind and heat the corona \citep[see e.g.,][]{Aschwanden:05}. The wind and magnetic field
solutions evolve together, allowing the magnetic field topology to influence the wind appropriately, 
as observed in the case of the solar wind in the heliosphere \citep{Phillips.etal:95, McComas.etal:07}.

\subsection{Simulations}
\label{sec:Simulations}

\begin{figure}[h]
\center
\includegraphics[trim = 1.3in 5.in 
.1in 1.4in,clip,width = 4in]{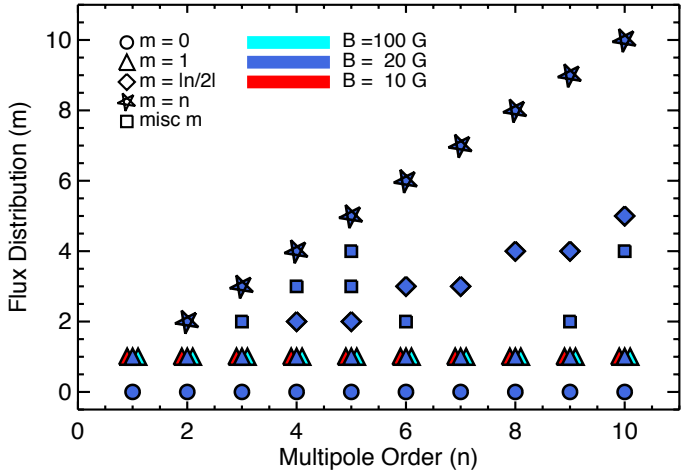} 
\caption{Grid of simulations of magnetic multipole moment $n$ (x-axis) and flux distribution $m$ (y-axis). The $m=1$ case for all $n$s are taken from CG15. The $33$ new cases are illustrated with circles for $m=0$, diamonds for $m=|n/2|$, stars for $m=n$, and squares represent other values of $m$. All the new cases are assuming a magnetic field strength of $20~G$. } 
\label{fig:nm}
\end{figure}

IN CG15 we investigated the mass and angular momentum loss rates for different levels of complexity of the magnetic field ($n$).  In order to study the role of the distribution of flux ($m$) for a given complexity ($n$), we perform three-dimensional stellar wind simulations for a hypothetical solar-mass star 
with a solar rotation period ($\sim 25$ days) for peak field flux densities of $B=20$~G and for different magnetic topologies.  It is arguable whether the magnetic field strength or the magnetic flux should be kept constant when comparing wind results for different field topologies ($n$).
However, we have shown in CG15 that keeping the flux or the strength constant makes no difference when studying the trends in behavior of mass and angular momentum losses with magnetic complexity.  It is worth noticing that magnetic flux only depends on $n$ and on the field strength, and it is independent of $m$ (see CG15 for details). Therefore, for a given $n$ both flux and field amplitude are constant across different values of $m$.  Our grid of fiducial magnetograms consists of the different allowed spherical harmonic flux distributions of the ten first magnetic moments (labelled by $m= 0 \,-\, n$ ).  Each term in the multipolar expansion of a magnetic map is given by a spherical harmonic $Y^{m}_{n}(\theta, \psi)= N e^{im \psi} P^{m}_{n}(\cos \theta)$, where $n$ is the order in the multipolar expansion that is responsible for the complexity of a map (number of polarity change rings) and $m$ controls how the magnetic flux is distributed.  In CG15 we studied how the efficiency of mass and angular momentum loss rates depend on the complexity of the magnetic fields on the stellar surface.  Here we extend this by studying the effect of different distributions of that magnetic flux (different $m$s for a given $n$).  In CG15 we simulated the first $10$ orders of the multipolar expansion ($n=0-10$) with $m=1$ for three different magnetic field strength ($10~G$, $20~G$, and $100~G$). Here we complete that sample by exploring the different $m$s in for all the $n$s ($1-10$) in the $20~G$ case.  For each $n$ we simulate the solutions with $m=0, 1, n/2, n$, plus other random cases (see Figure~\ref{fig:nm} where our grid of simulations is illustrated). This adds $33$ simulations to the $40$ performed in CG15.  As in CG15, the phase term becomes irrelevant for pure modes in a sphere because it represent a shift in azimuthal angle. 

\begin{figure*}[h]
\resizebox{\hsize}{!}{\includegraphics[trim = 0.3in 0.3in
  0.15in 2.85in,clip, width = 2.5in]{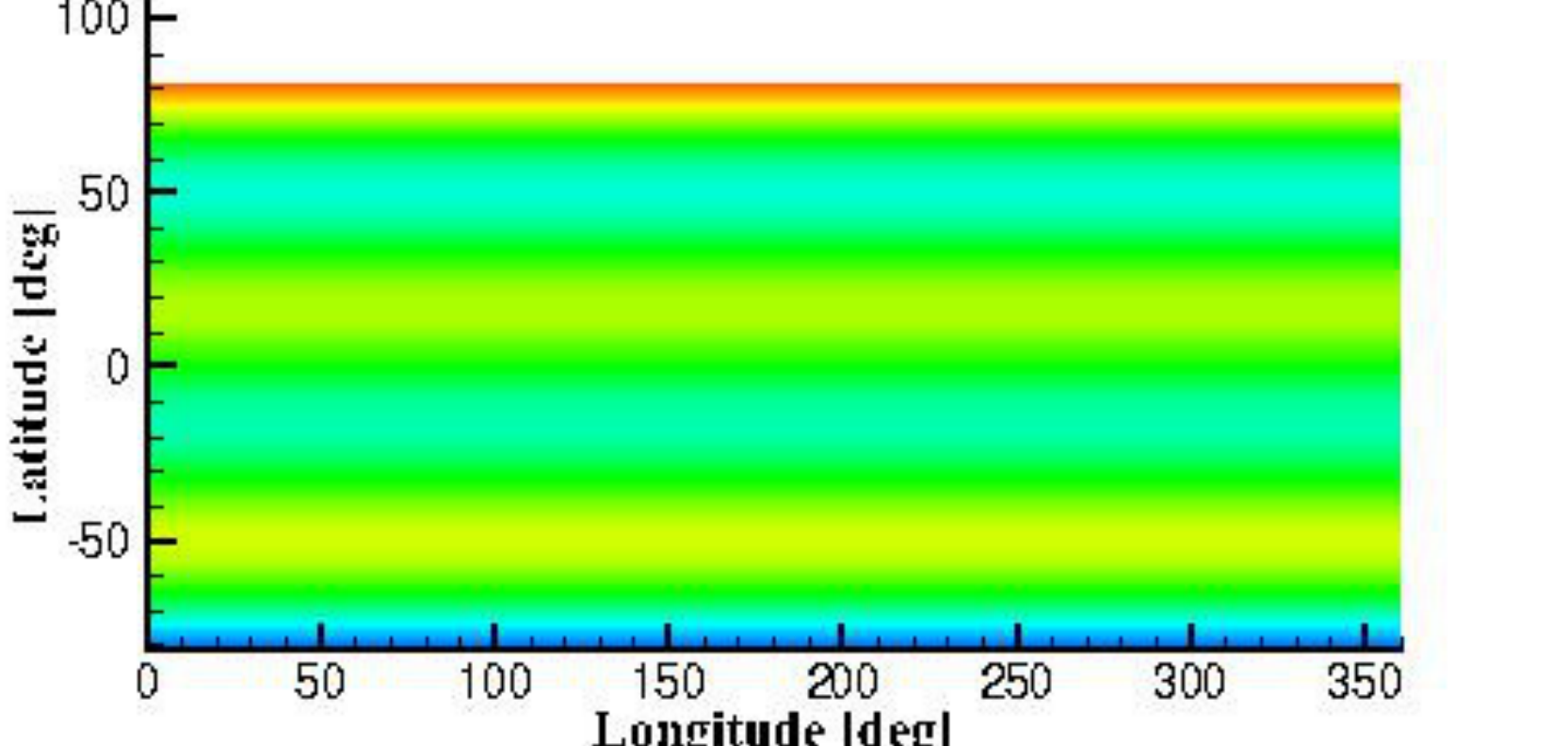} 
\includegraphics[trim = 0.3in 0.3in
  0.15in 2.85in, clip, width =2.5in]{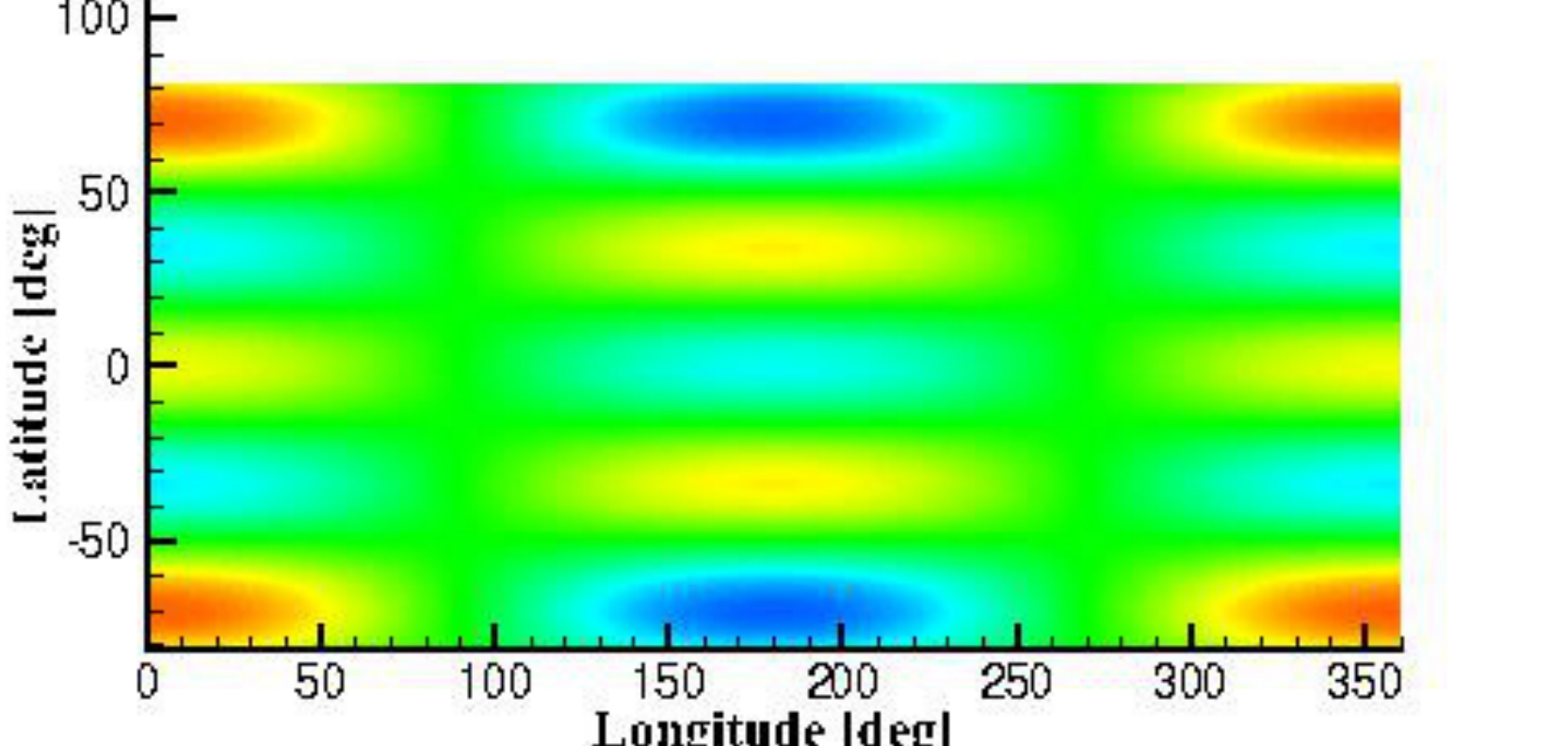} 
  \includegraphics[trim = 0.3in 0.3in
  0.15in 2.85in,clip, width = 2.5in]{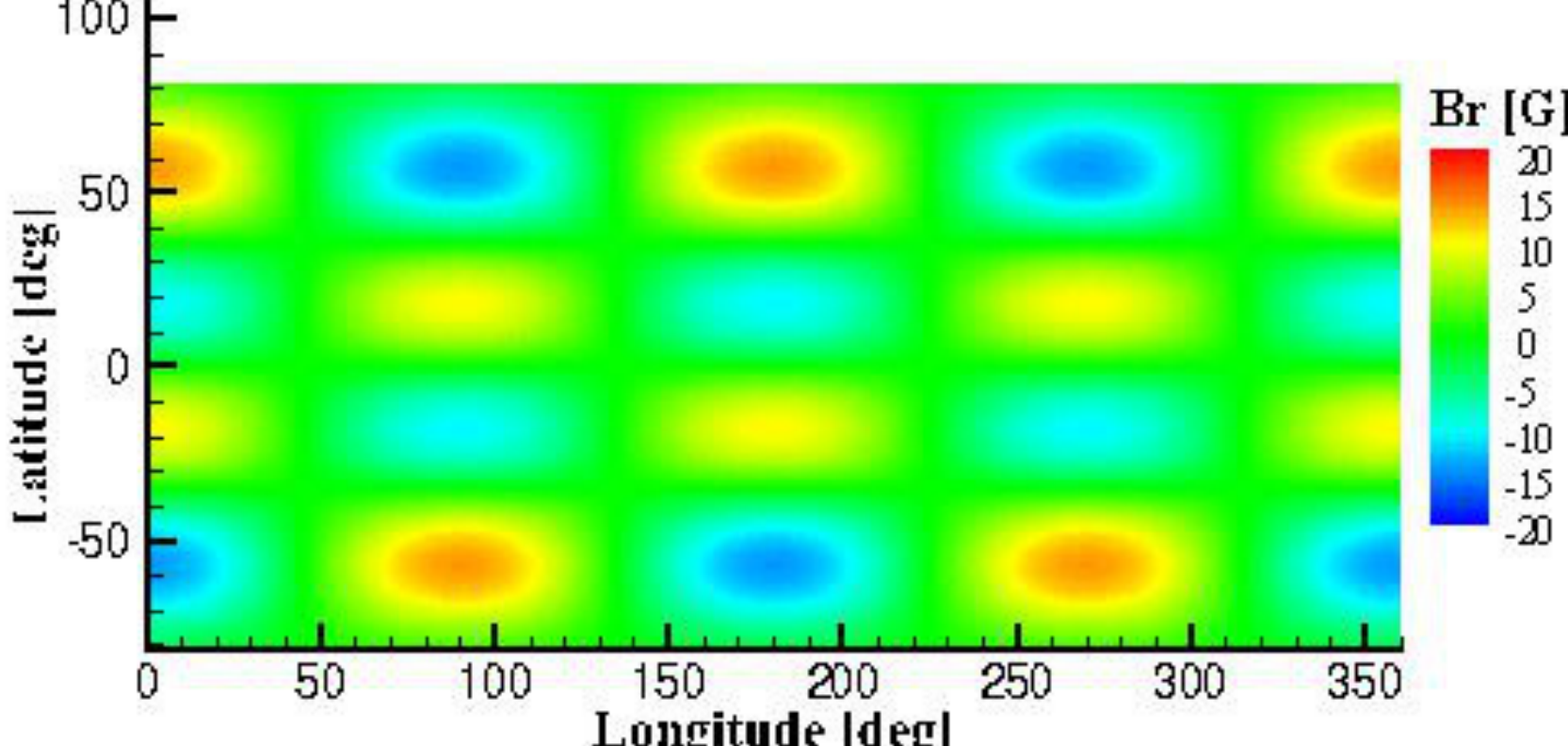} }
\resizebox{\hsize}{!}{\includegraphics[trim = 0.3in 0.3in
  0.15in 2.85in,clip, width = 2.5in]{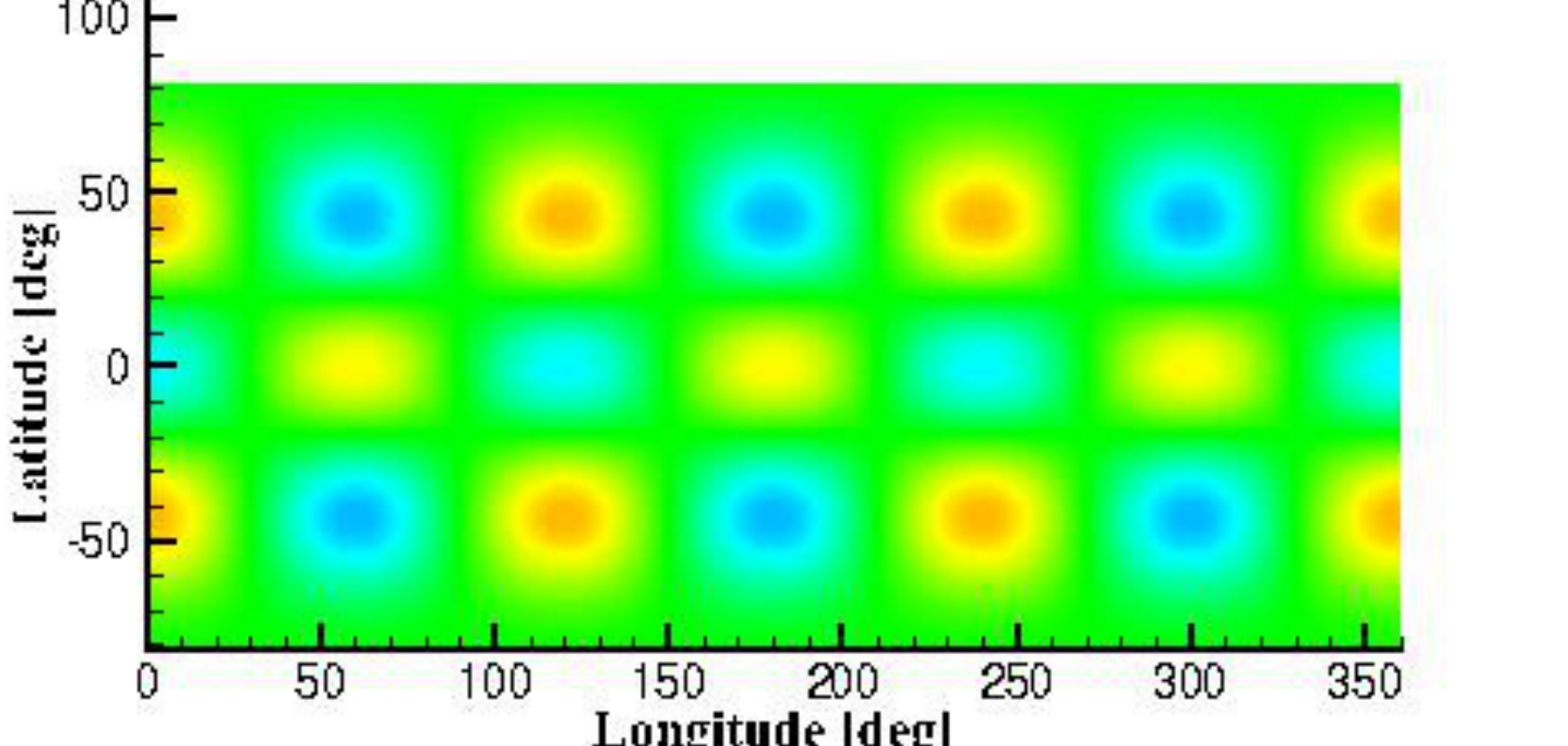} 
\includegraphics[trim = 0.3in 0.3in
  0.15in 2.85in,clip, width = 2.5in]{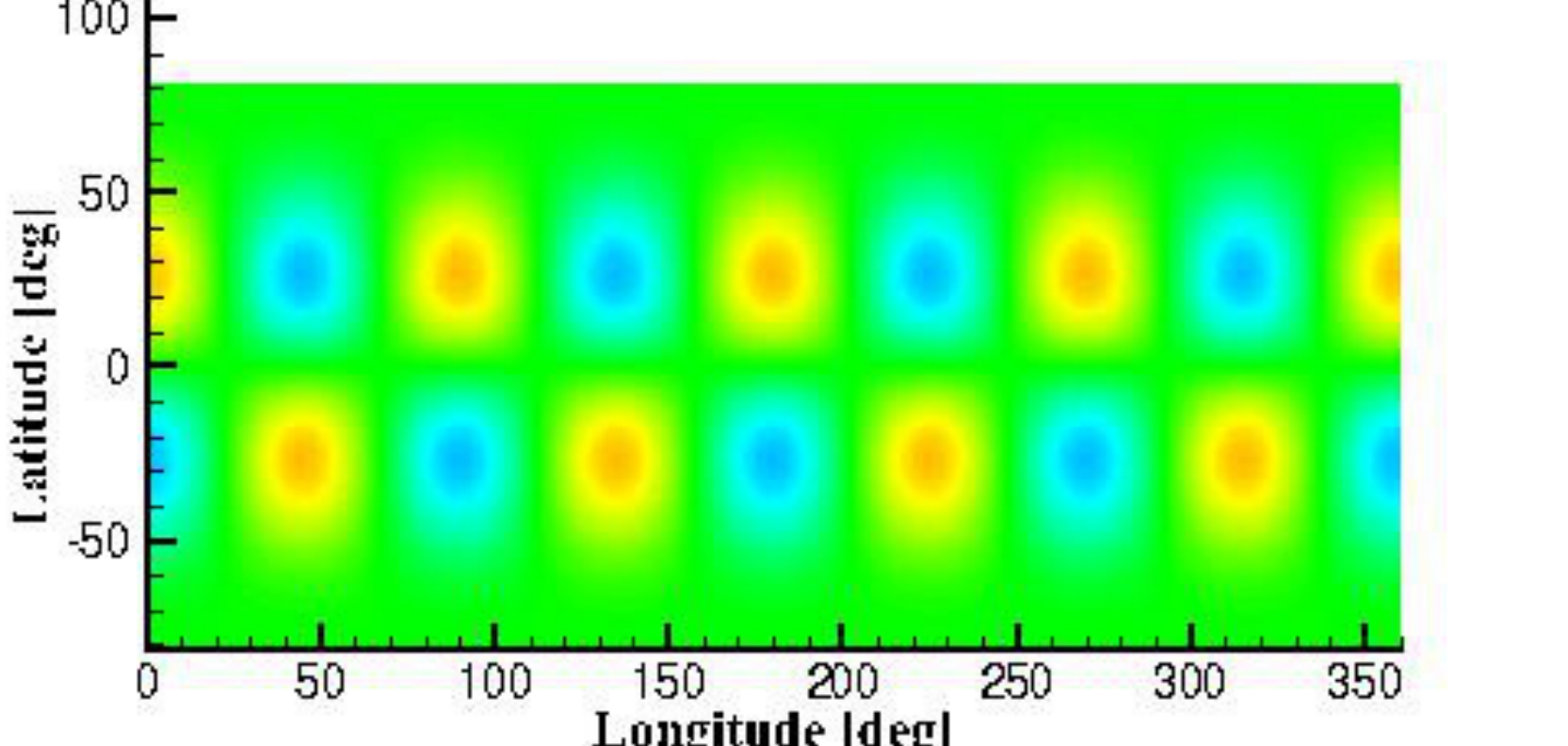} 
 \includegraphics[trim = 0.3in 0.3in
  0.15in 2.85in,clip, width = 2.5in]{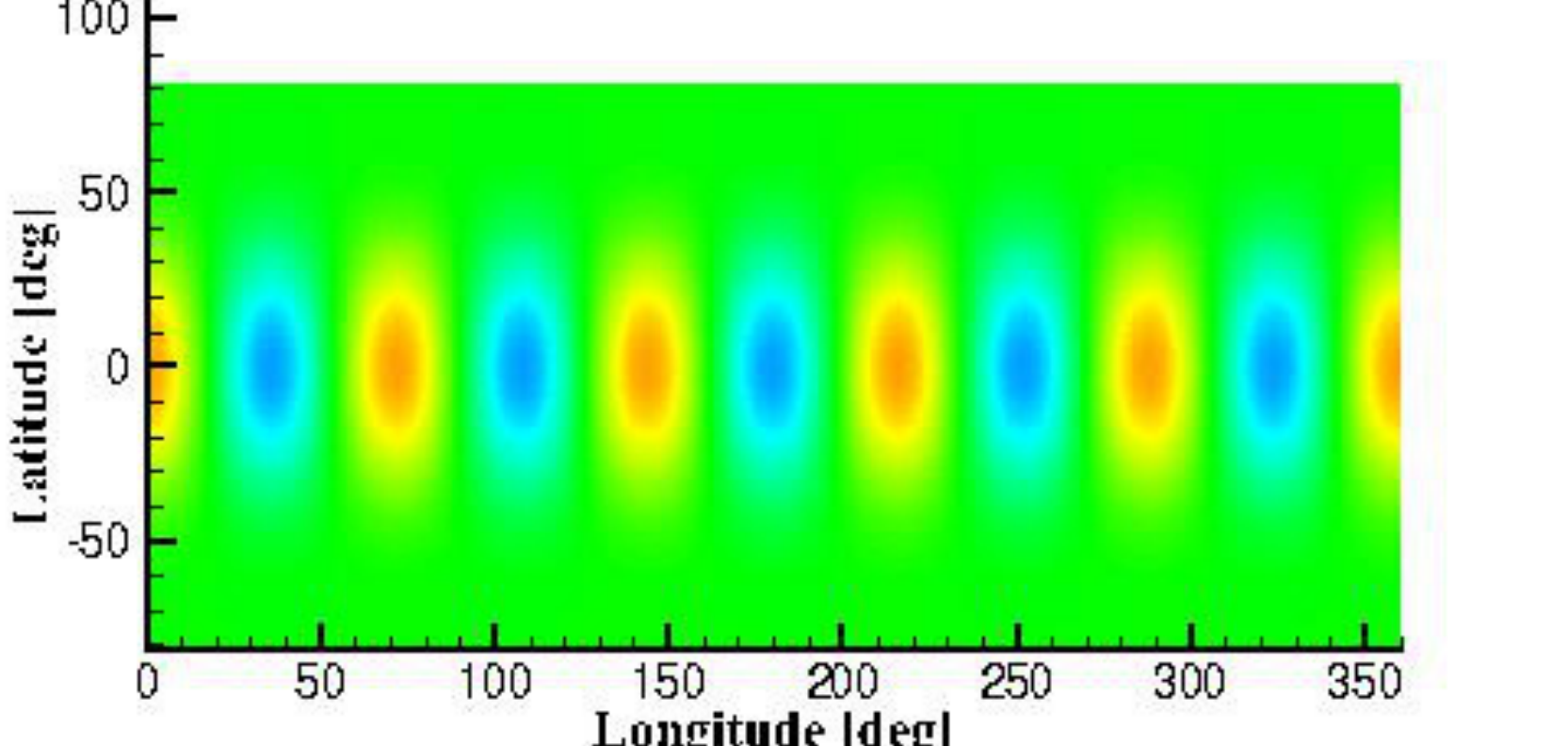} }
 \caption{Fiducial magnetograms for magnetic flux densities of increasing $m$  (from top left to right bottom) of a 5th order magnetic multipole
  $n=5$ and a 20~G amplitude field. }
\label{fig:magnetograms}
\end{figure*}

An example of these fiducial magnetic maps for a fixed level of complexity ($n=5$) is shown in Figure~\ref{fig:magnetograms}.  

As in CG15, from the three-dimensional model solutions we extract the wind
density, $\rho$, and speed, $\mathbf{u}$, over the
Alfv\'en surface and at the stellar surface.   The Alfv\'en surface itself is determined by finding the surface for which the wind speed reaches the local Alfv\'en speed, $v_A=B/\sqrt{4 \pi \rho}$,
neglecting the contribution of the electrons to the mass density, $\rho$, and for a pure hydrogen wind.  We then compute
the mass and angular momentum loss rates at each point on the Alfv\'en
Surface (see CG15 for details).
We also calculate the amount of open magnetic flux through
a spherical surface outside of the Alfv\'en surface.  

Having explored both parameters involved in the spherical harmonic decomposition (and based on $73$ 3D state-of-the-art numerical simulations) we deduce analytical expressions to estimate the mass and angular momentum losses from the flux and complexity of the magnetic map (see \ref{sec:morphterm}).  In addition, and in order to test these scaling laws, we perform simulations for $8$ real magnetograms (see Figure~\ref{fig:realmagnetograms}), including solar maximum and solar minimum, and compute their mass and angular momentum loss rates.  For our sample of real stars, we decompose the magnetograms in spherical harmonics ($n= 1 - 6$ for stars and $n= 1 - 10$ for the sun) and compute the magnetic flux-weighted average of $n$ as a complexity proxy, called $n_{av}$.  It was shown by \cite{Garraffo.etal:13} that magnetic flux stored in low latitude spots ($\le 45$~deg) does not alter the mass and angular momentum loss rates. For that reason we only consider the large scale morphology in this paper and we do not include the spots in the calculation of total flux for the Sun, which turns out to be  $1.86 \cdot 10^{23}$~Mx for solar minimum and $5.87 \cdot 10^{23}$~Mx for solar maximum, which are consistent with the observed values (see, for example,  \citealt{Jin.Wang:14, Wang.etal:09} ).   We then obtain the expected mass and angular momentum loss rates using our analytical expressions, and compare the expected rates with the ones obtained from the simulations.

\section{RESULTS}
\label{sec:Results}

\subsection{Flux Distribution Dependence}

\begin{figure*}[h]
\center
\includegraphics[trim = .1in 4.4in 
.1in 1.3in,clip,width = 5.1in]{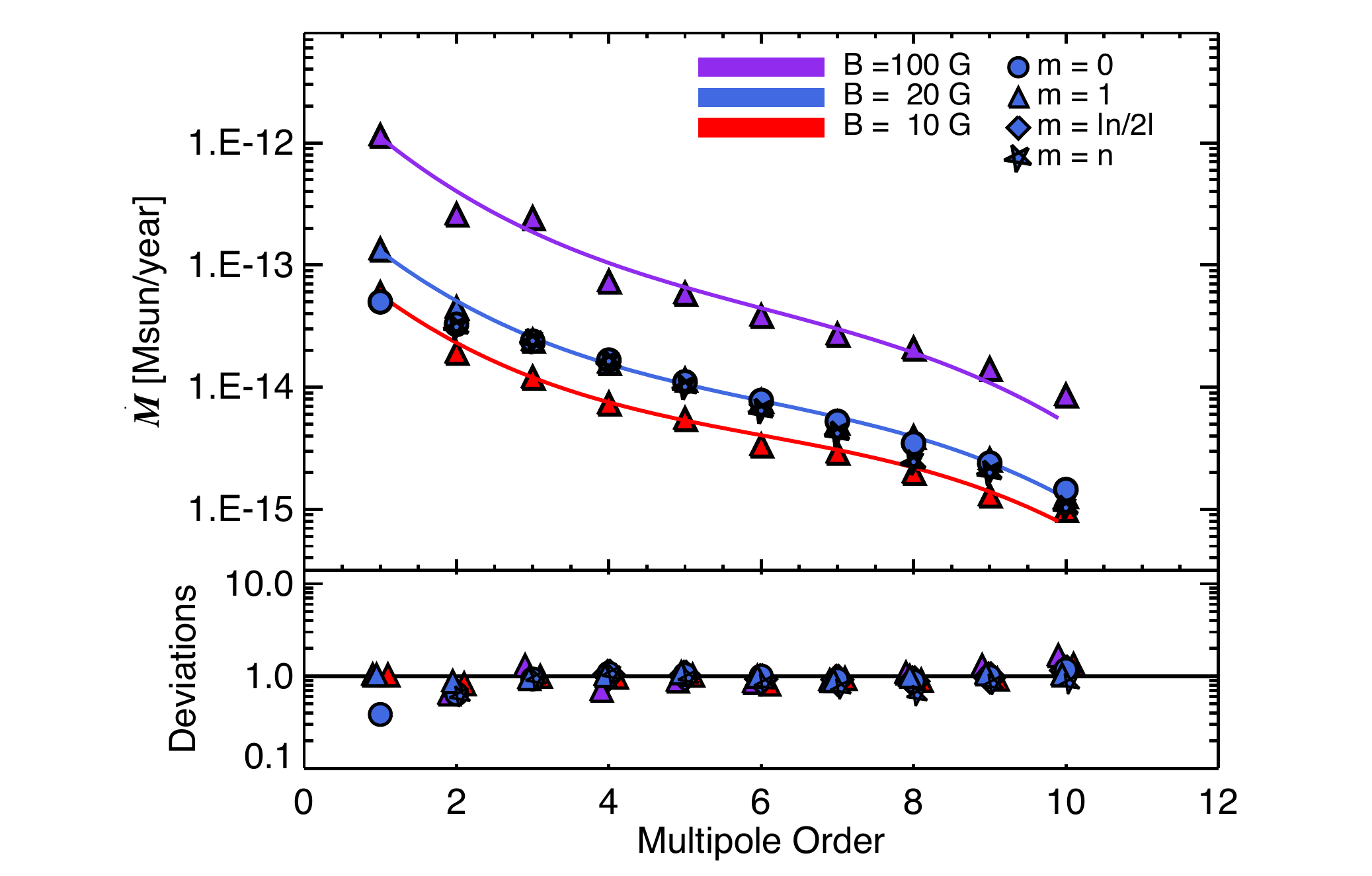}\\
\includegraphics[trim = .1in 4.4in 
.1in 1.3in,clip,width = 5.1in]{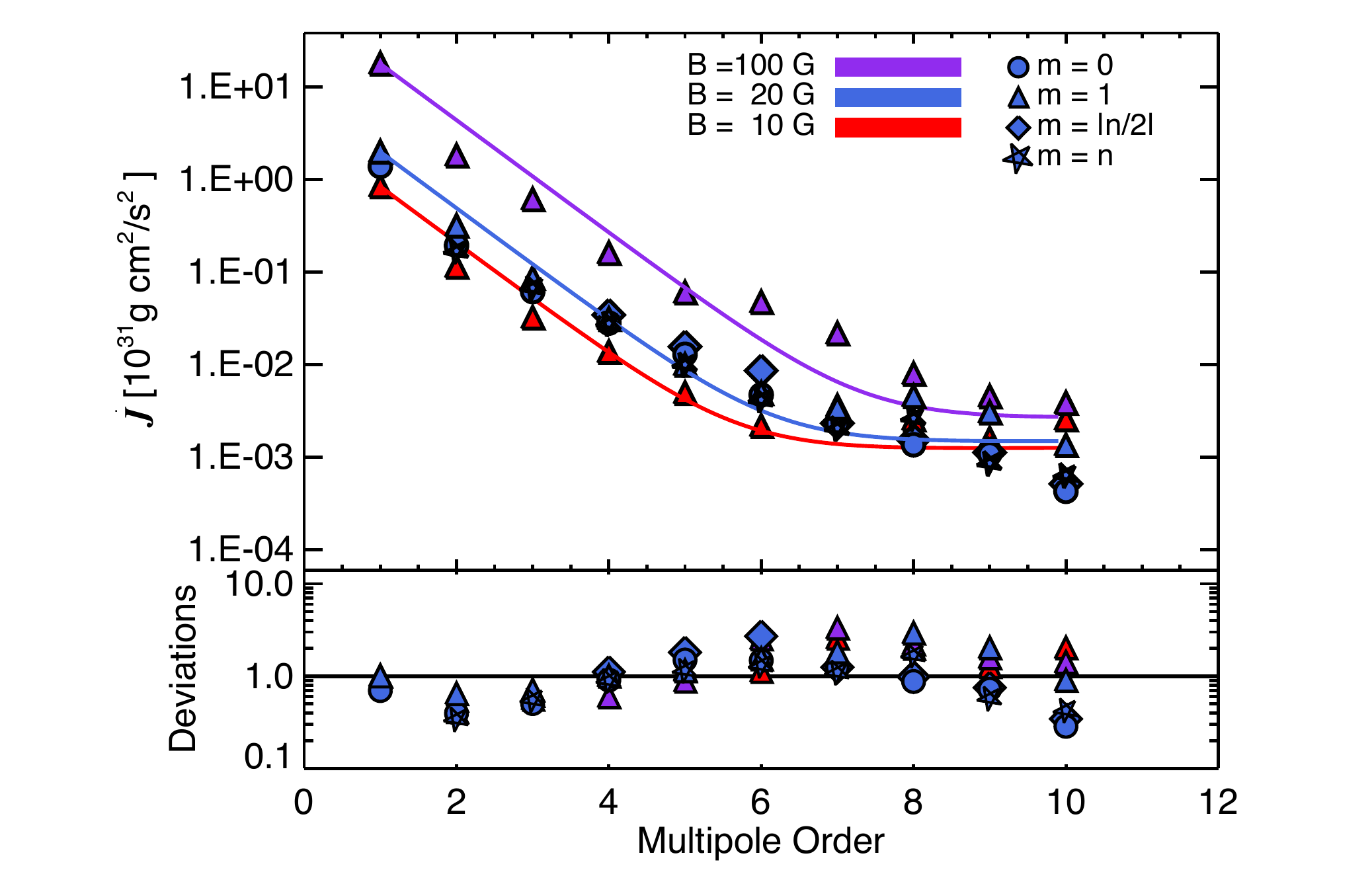}\\
\includegraphics[trim = .1in 4.4in 
.1in 1.3in,clip,width = 5.1in]{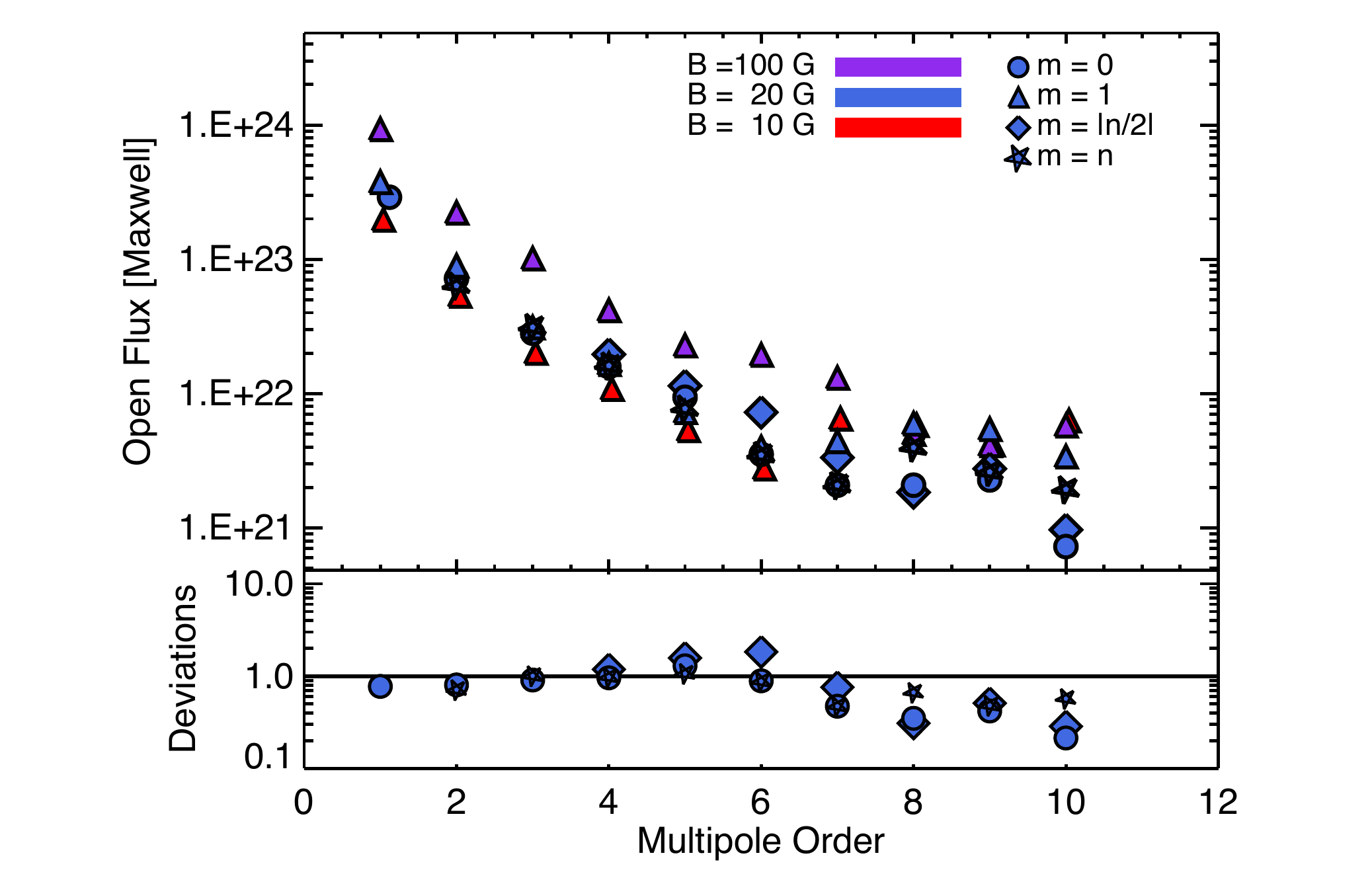}
\caption{Mass ($\dot M$ top), angular momentum ($\dot J$ middle) loss
  rates, and open flux (bottom) for different levels of complexity $n$ (x-axis), different field strengths (10~G in red, 20~G in blue, and 100~G in cyan), and different distributions of magnetic flux with different symbols (only for the 20~G case).  The smooth curves are the analytical relations derived in Section~\ref{sec:morphterm}. The bottom plot in the two top panels shows the logarithmic deviations of the simulations over the analytical curves, while the bottom panel shows the logarithmic difference between the open flux of the $m=1$ case and the rest, with the same symbols and colors.} 
\label{fig:aml}
\end{figure*}

The mass and angular momentum loss rates, and open flux computed for each case in our grid of models are plotted in Figure~\ref{fig:aml}.  The triangles are the simulations reported in CG15, which correspond to the $m=1$ case for $n=0 - 10$.  We left the remaining cases out of this plot (shown as squares in Figure~\ref{fig:nm}) for the sake of clarity, but the results obtained from those are consistent with the ones shown here.  \\

Each term in the multipolar expansion, labeled with $n$, represents a different complexity with a fixed number of rings in which polarity changes (see Figure~\ref{fig:magnetograms}).  From our results it is clear that the level of complexity $n$ is the dominant morphology factor determining mass loss rates and the resulting angular momentum loss rates.  It is also shown that the good correlation between angular momentum loss rate and open flux discussed by \cite{Garraffo.etal:15} and \cite{Reville.etal:15a} still holds.

Analytical approximations to the loss rates are derived in Section~\ref{sec:morphterm} below and are also illustrated in in Figure~\ref{fig:aml}.  The bottom plot in each panel shows logarithmic deviations with respect to the analytical model for the mass and angular momentum loss rates, and with respect to the case $m=1$ for the open flux.

\begin{figure*}
\center
\resizebox{\hsize}{!}{\includegraphics[trim = .1in .2in 
.1in .1in,clip, width =1.35in]{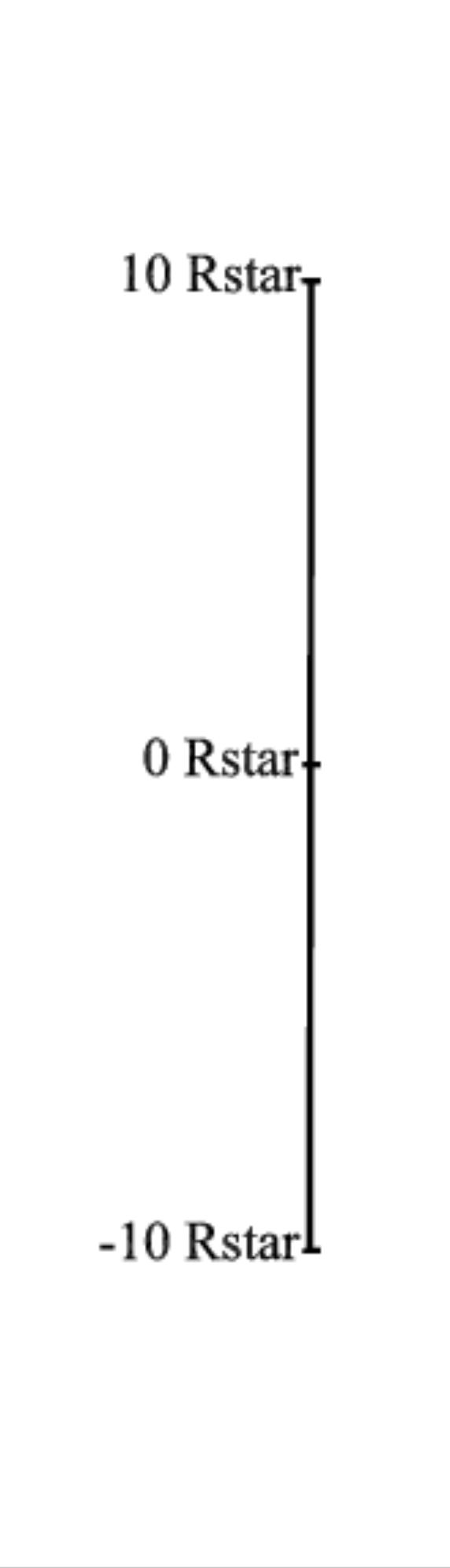}
\includegraphics[trim = .2in .3in 
.1in .1in,clip, width = 3.1in]{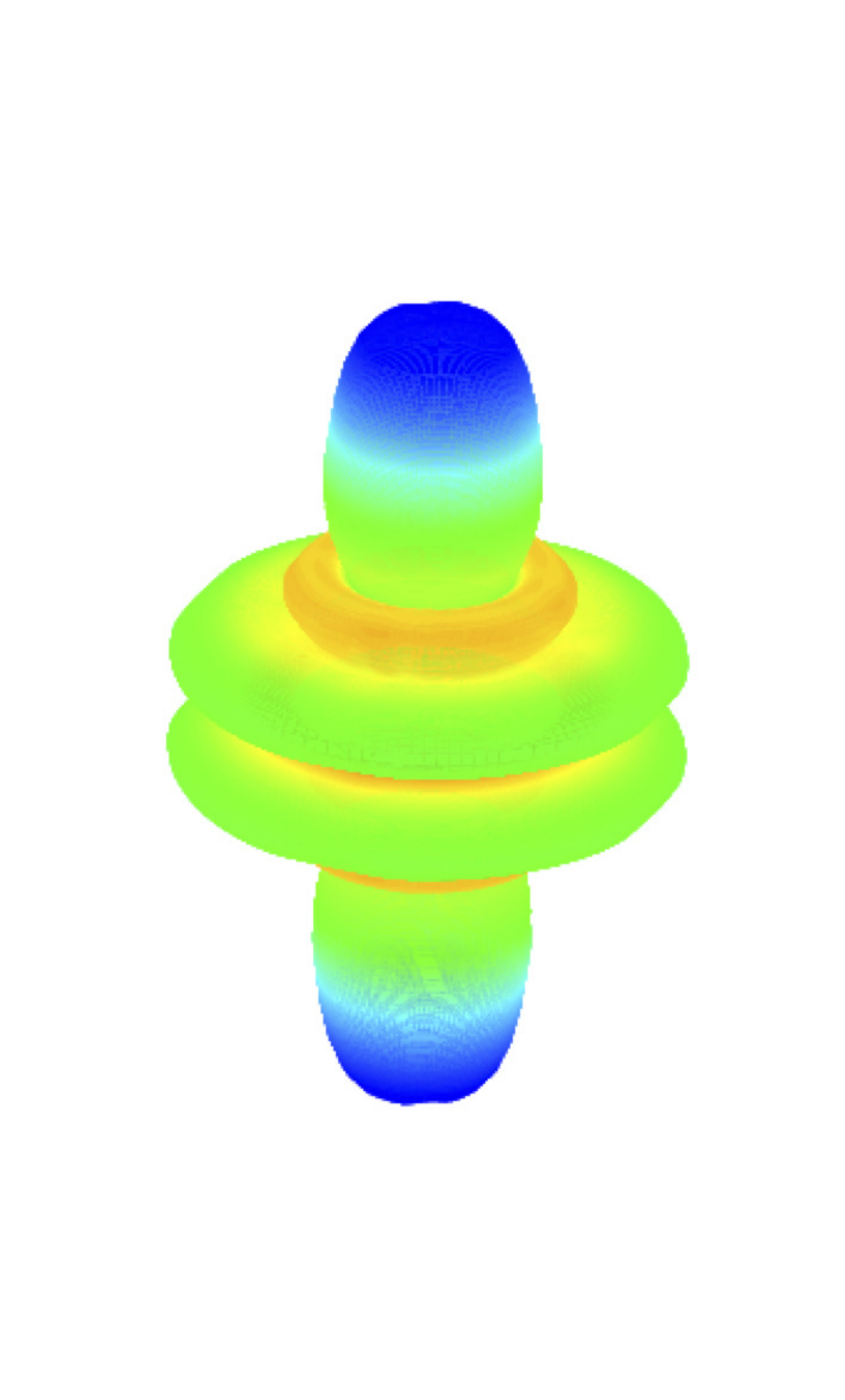} 
\includegraphics[trim = .2in .3in 
.15in .1in,clip, width = 3.1in]{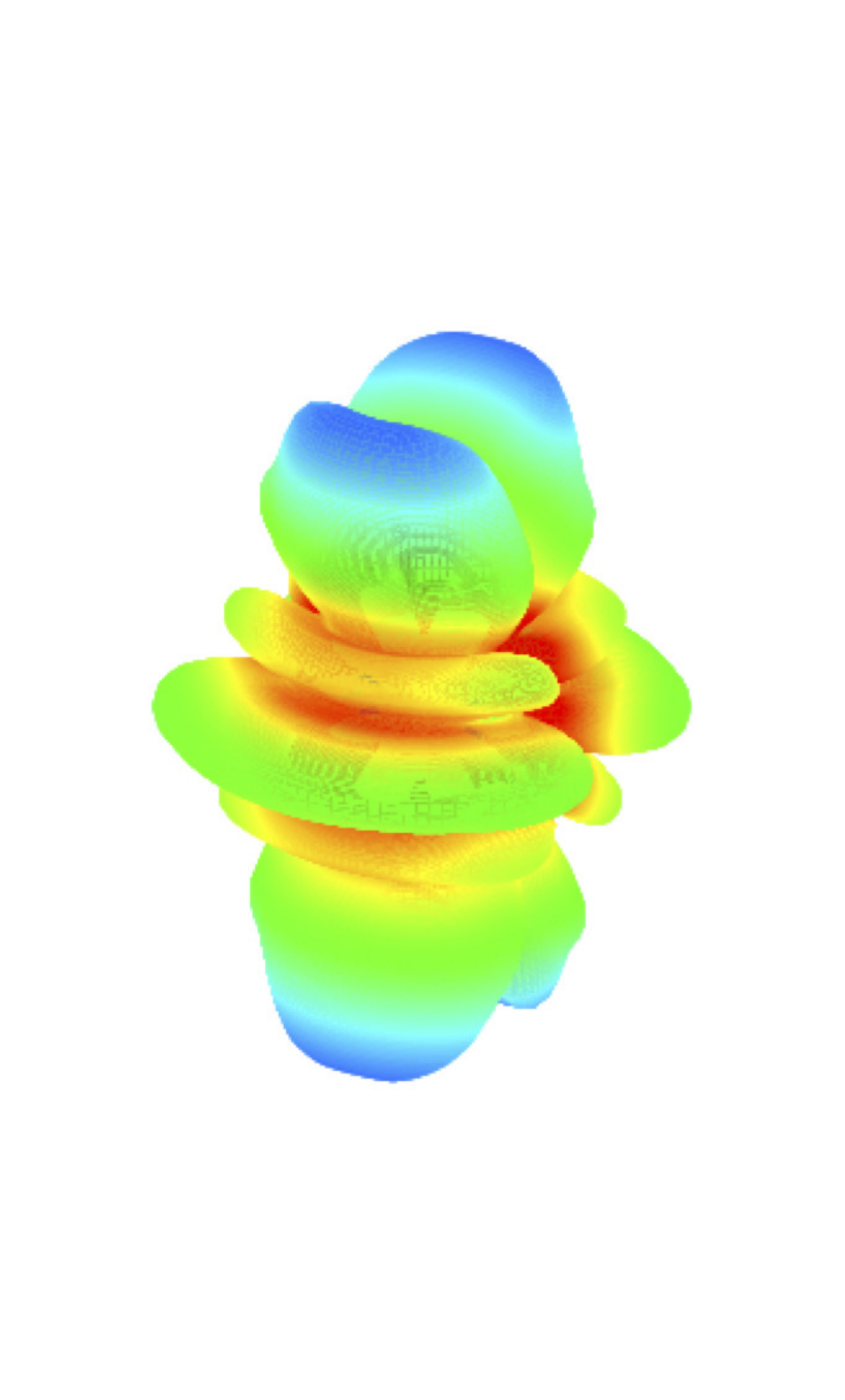}
\includegraphics[trim = .3in .3in 
.15in .02in,clip, width = 3.in]{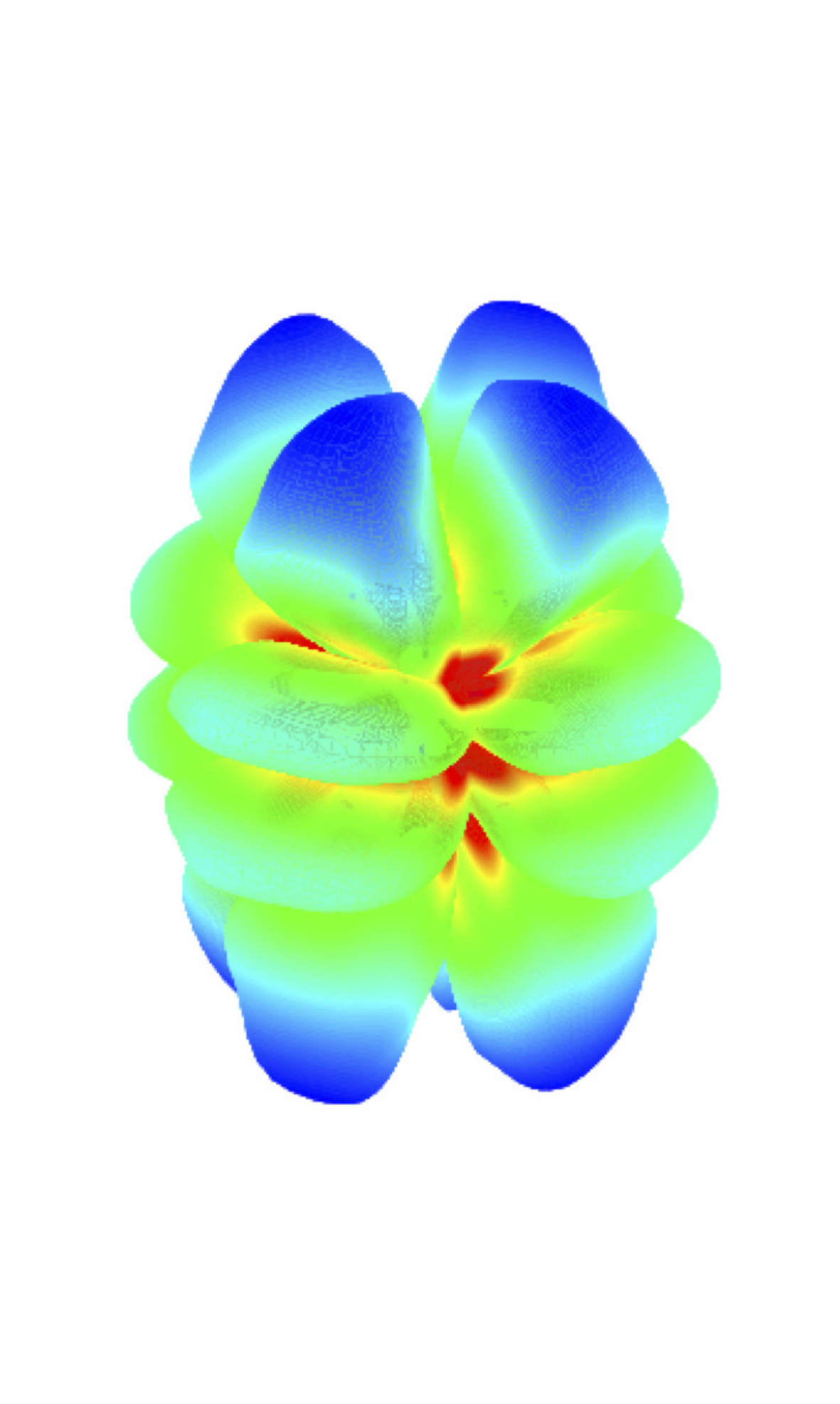}
\includegraphics[trim = .2in .3in 
.2in .15in,clip, width =3.1in]{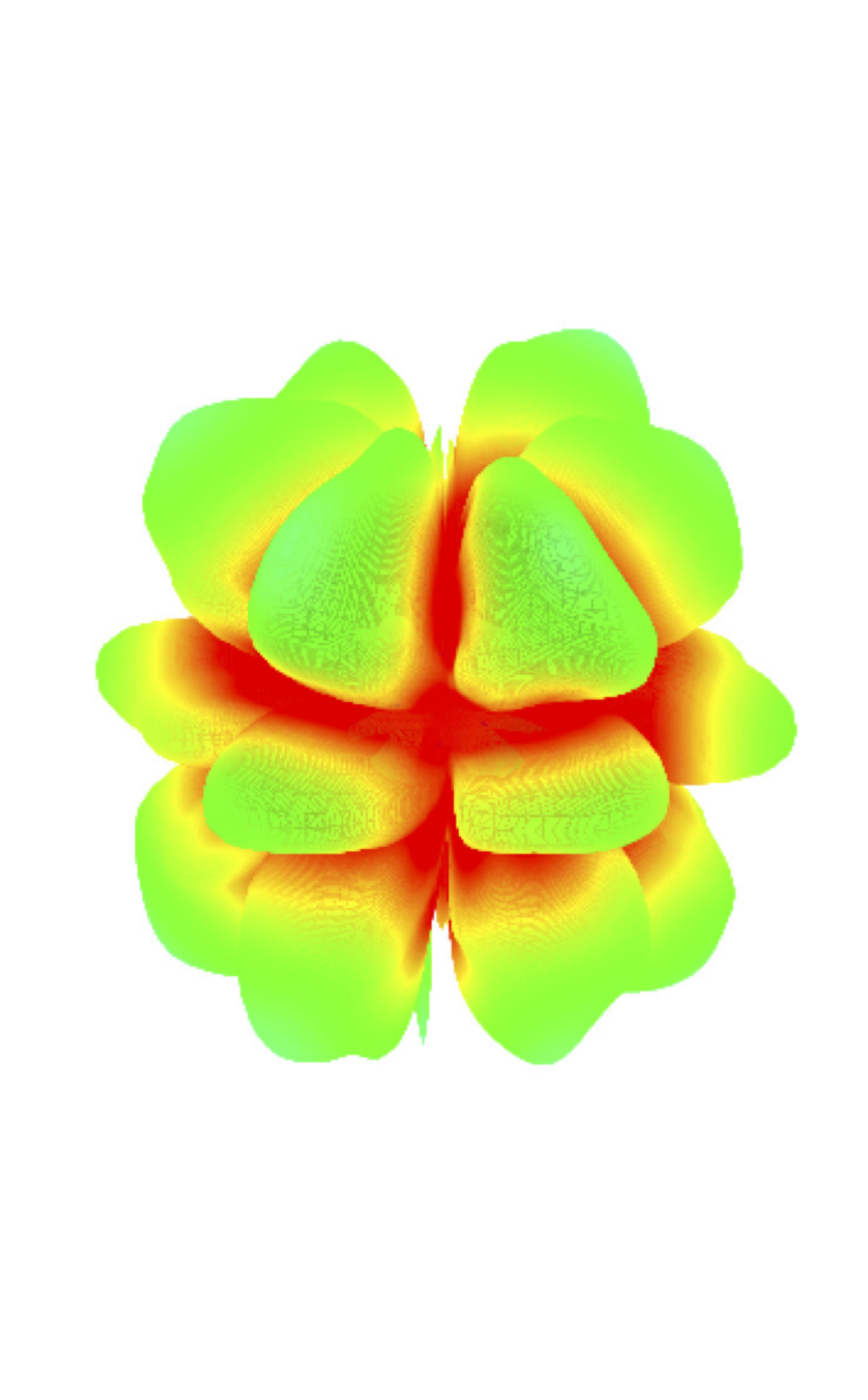}
\includegraphics[trim = .2in .3in 
.2in .15in,clip, width =3.1in]{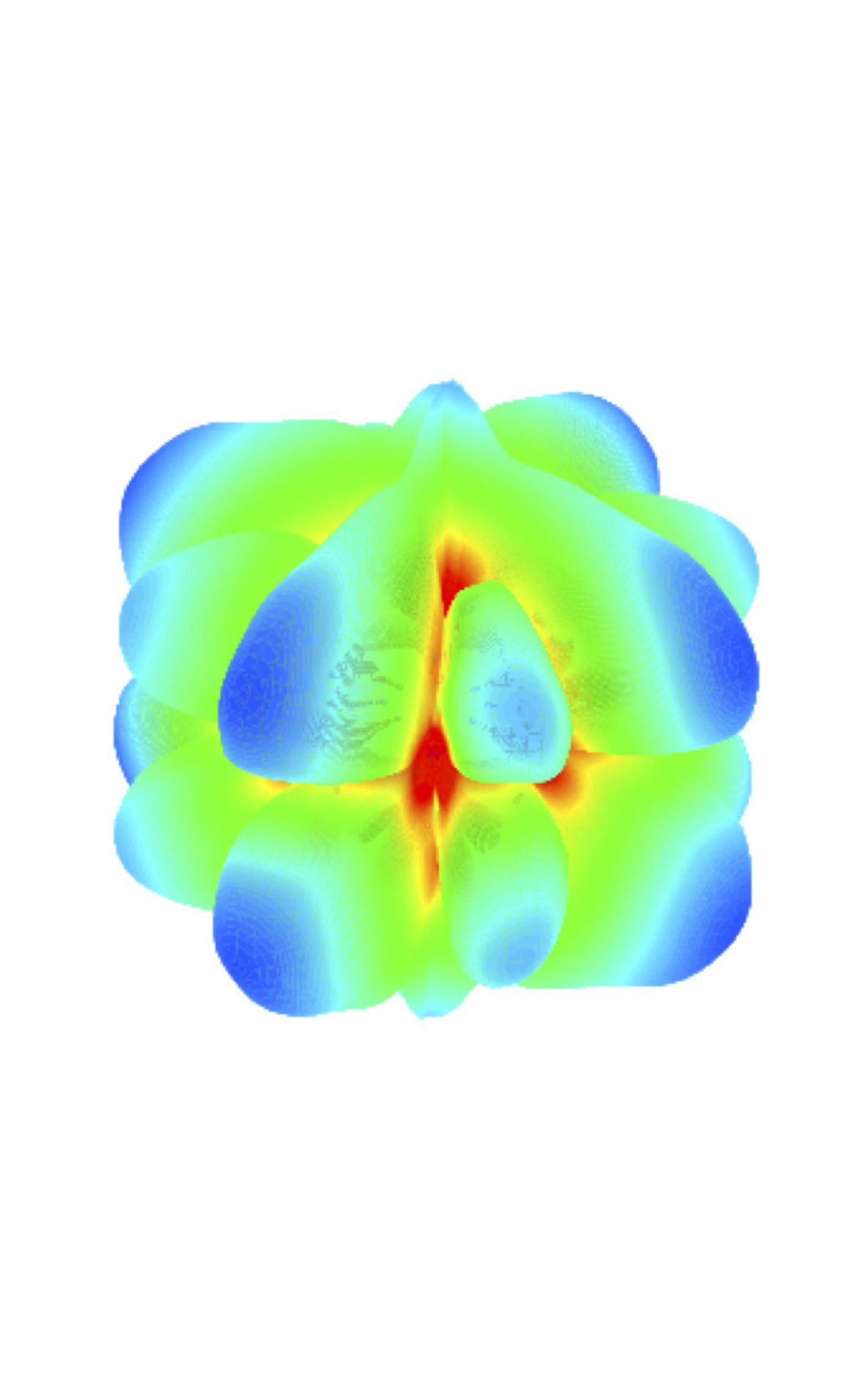}
\includegraphics[trim = .2in .2in 
.2in .15in,clip, width =3.1in]{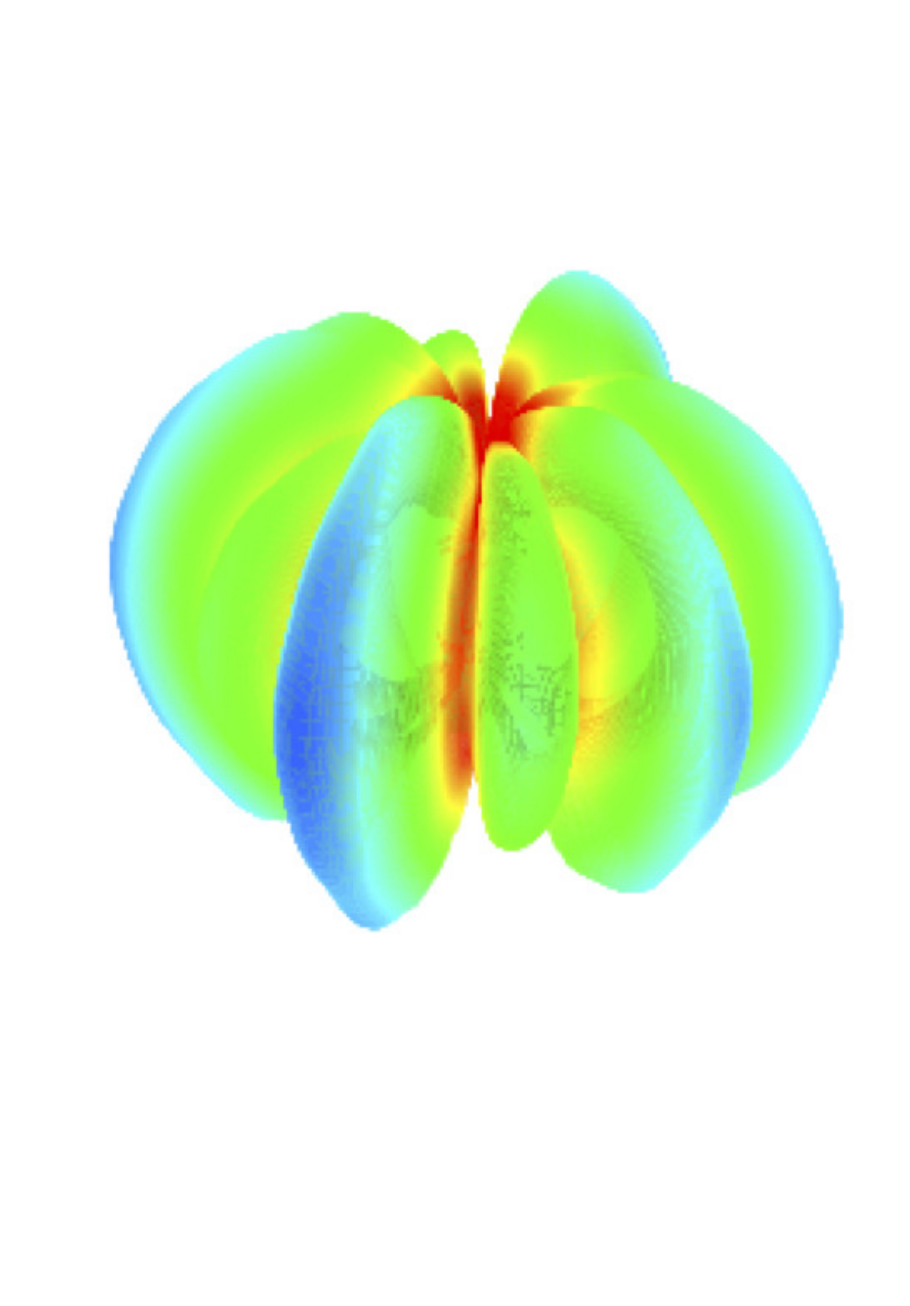}
\includegraphics[trim = .1in .25in 
.3in .1in,clip, width =1.6in]{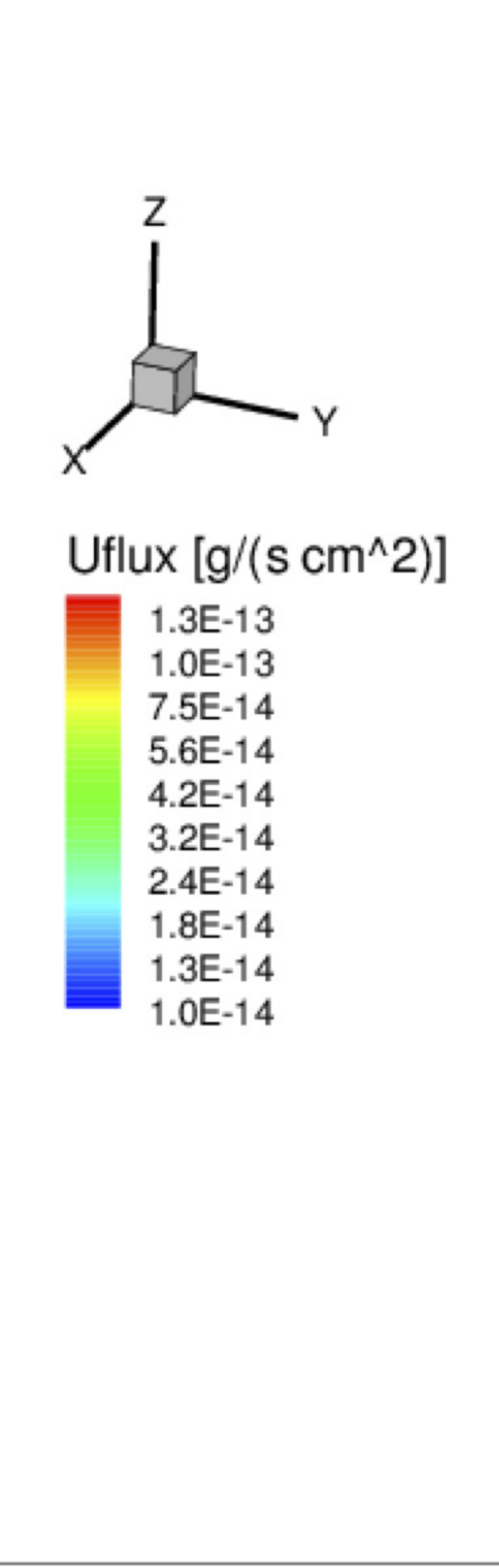}}
\resizebox{\hsize}{!}{\includegraphics[trim = .1in .4in 
.3in .1in,clip, width =.27in]{AS_Uflux_scale}
\includegraphics[trim = 1.75in .1in 
1.912in .7in,clip,width =.7in]{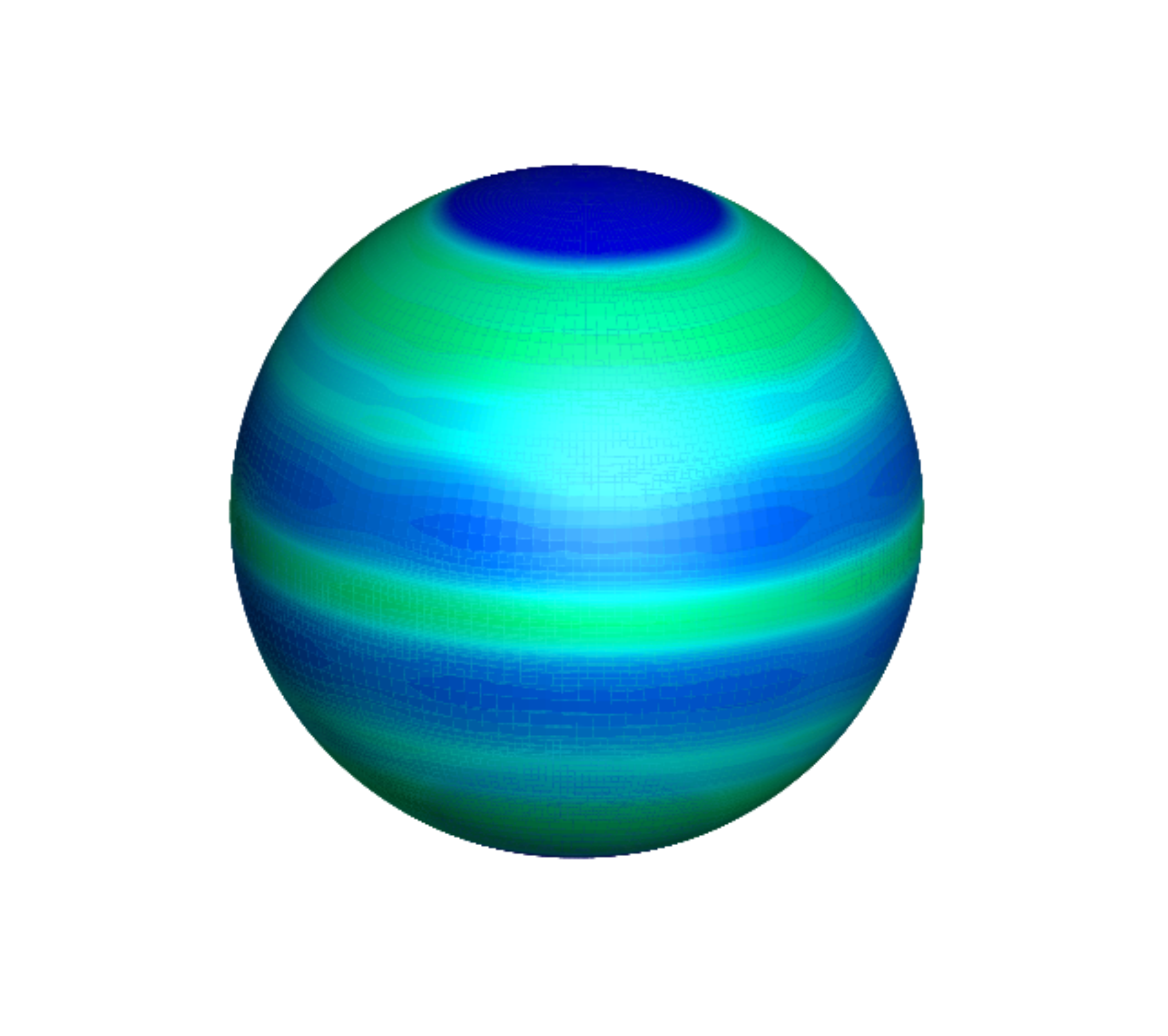} 
\includegraphics[trim = 1.75in .1in 
1.912in .7in,clip,width =.7in]{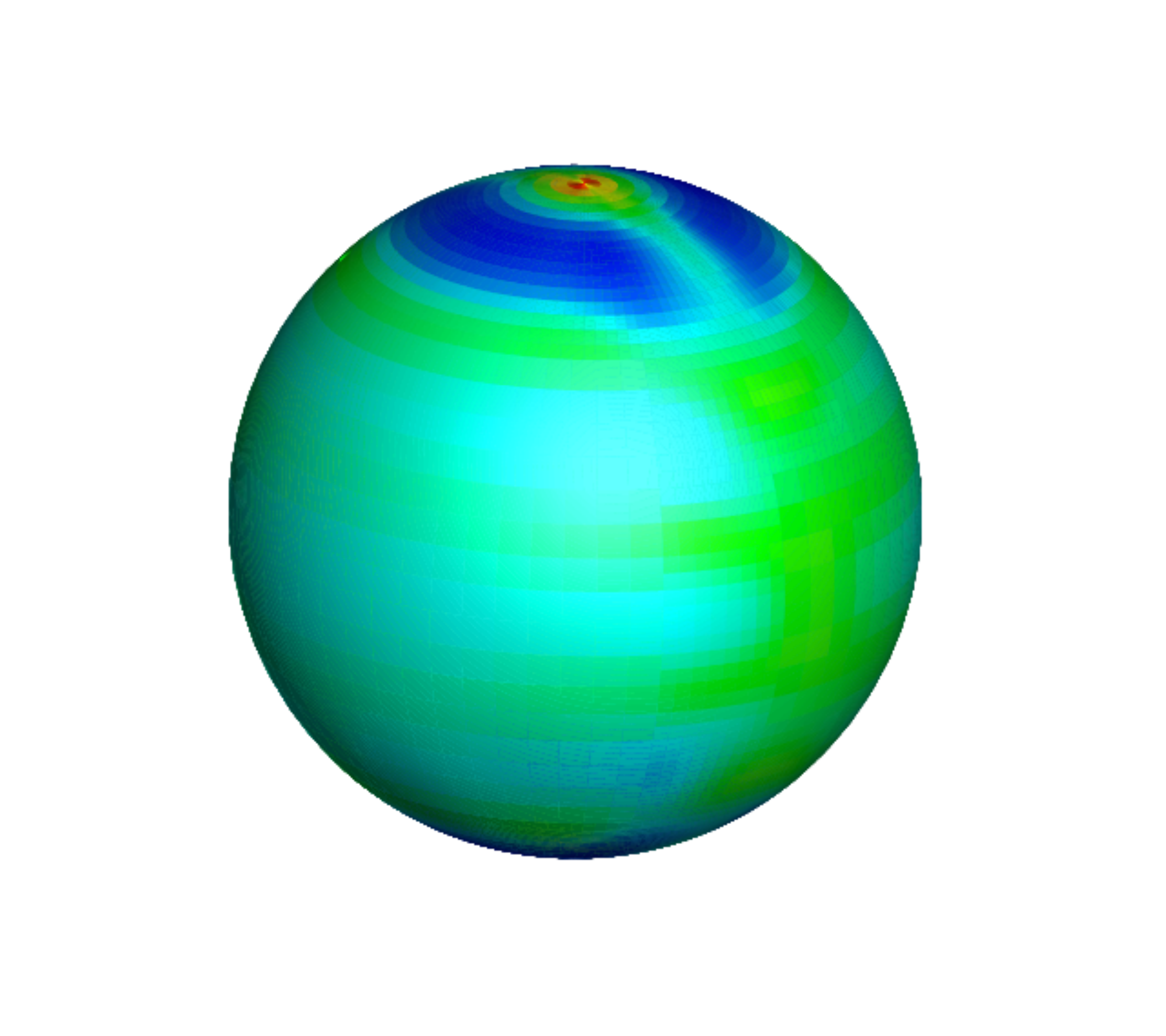}
\includegraphics[trim = 1.75in .1in 
1.912in .7in,clip, width = .7in]{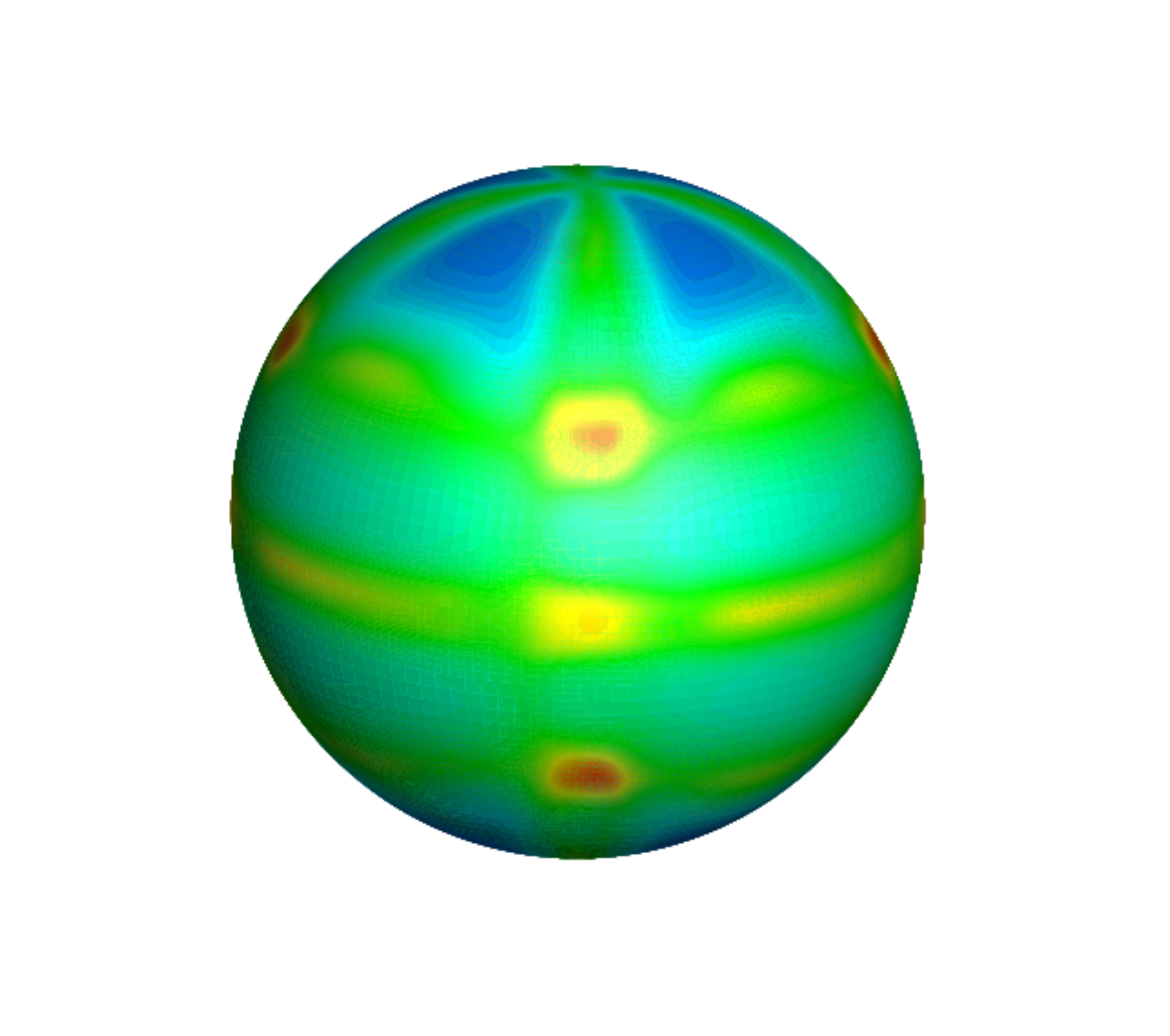}
\includegraphics[trim = 1.75in .1in 
1.912in .7in,clip, width =.7in]{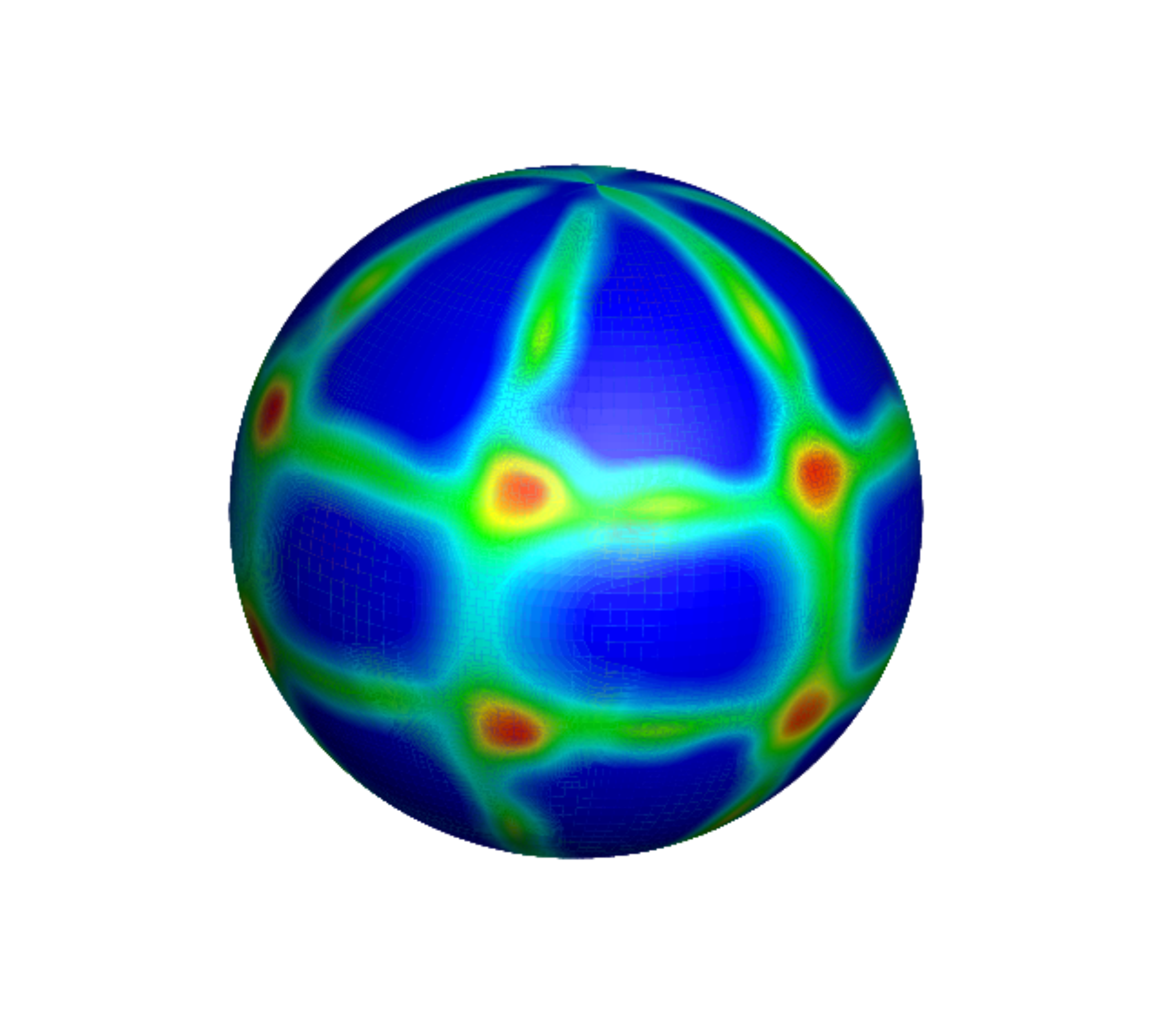}
\includegraphics[trim = 1.75in .1in 
1.912in .7in,clip, width =.7in]{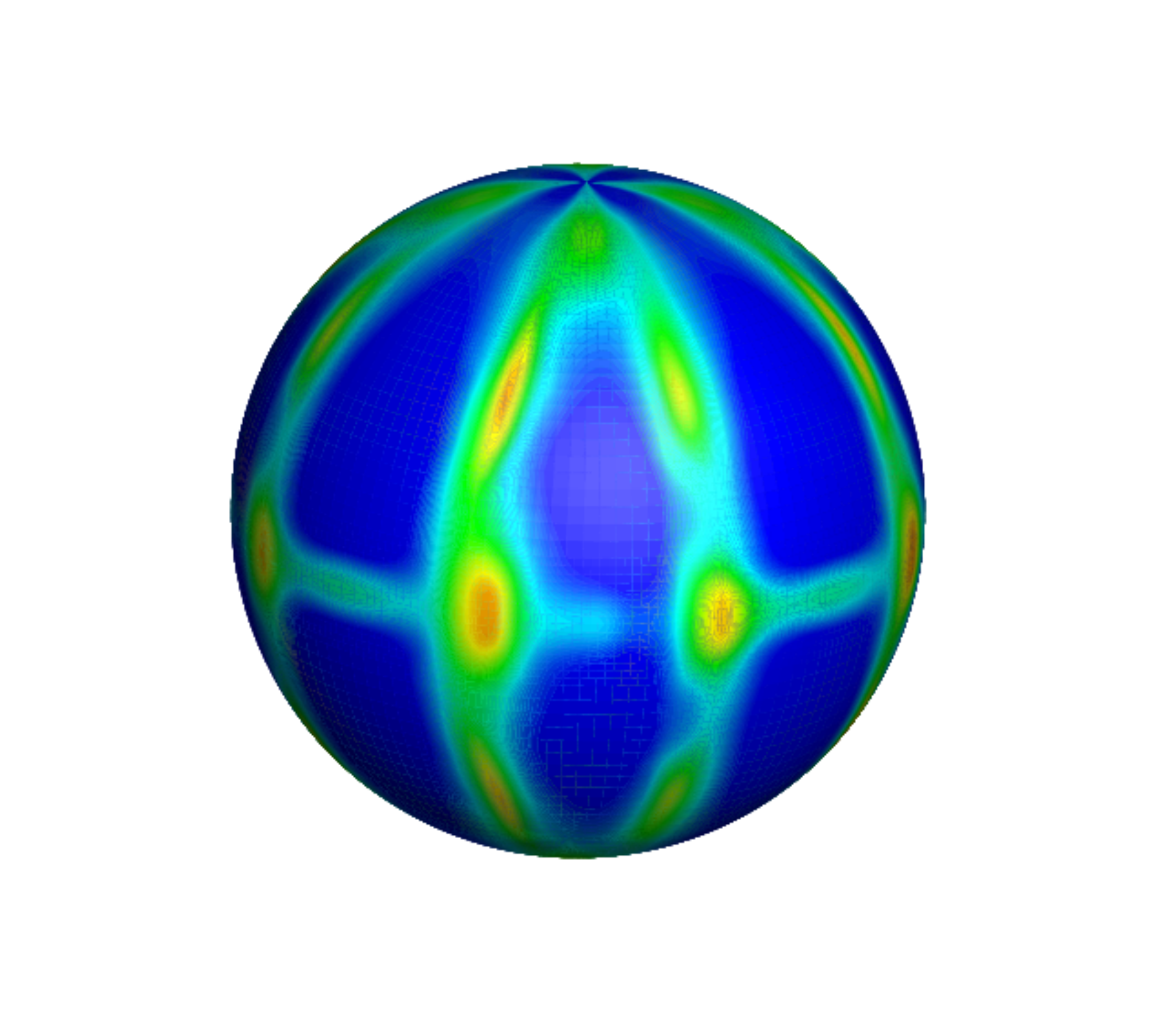}
\includegraphics[trim = 1.772in .1in 
1.89in .7in,clip, width =.7in]{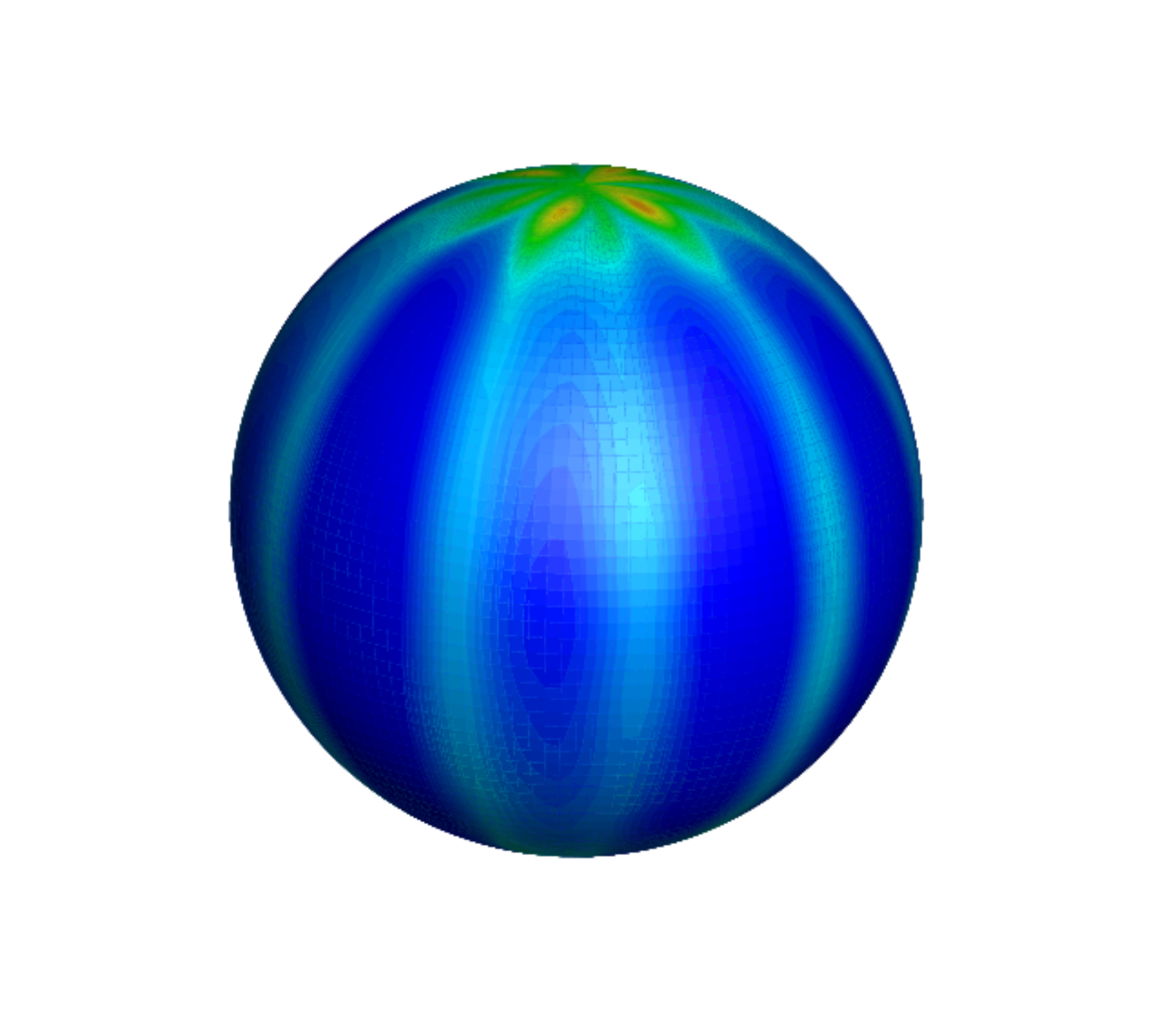}
\includegraphics[trim = .0in 1in 
.0in .0in,clip, width =.3in]{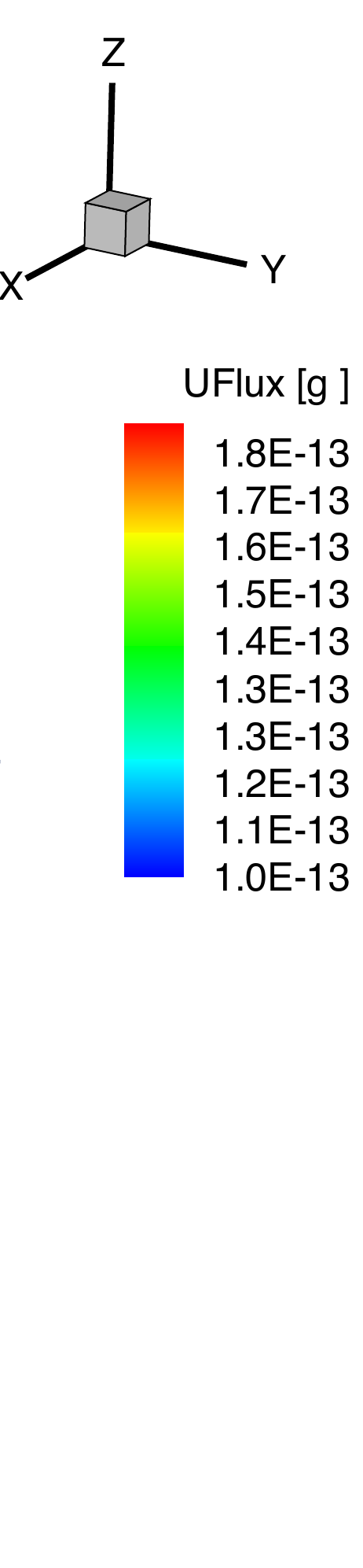}}
\caption{Top: Three dimensional Alfv\'en surfaces for 20~G fields of increasing 
$m$ (from left right bottom) of a 5th order magnetic multipole. Color scale represents the mass flux through it. Bottom: Mass flux through a $10R_{\sun}$ sphere.}
\label{fig:AS}
\end{figure*}

While the mass loss rates follow the smooth analytical trend quite precisely, there is some amount of dispersion (at each $n$) when it comes to angular momentum loss rates.  As we discussed in CG15, there are three interrelated aspects to the angular momentum loss: the mass flux, the Alfv\'en radius over which it acts as a rotational brake, and the
latitude at which the mass release happens.  The mass flux remains constant when changing $m$ (see top panel in Figure~\ref{fig:aml}). The radial dependence of the magnetic field which largely controls the Alfv\'en surface extent is proportional to
$1/r^{n+1}$, where $n$ is the magnetic multipole order, and is independent of $m$ (as discussed in CG15; see Figure~\ref{fig:AS}).
Therefore, the size of the Alfv\'en surface should be independent of $m$ too, as confirmed with our own simulations. We show that this is the case in Figure~\ref{fig:AS} (we only show one case for illustration, and we picked $n=5$ to be consistent with the choice of magnetograms shown in Section~\ref{sec:Simulations}, but this is true for all $n$).  However, as is clear from Figure~\ref{fig:AS}, most of the mass loss comes from regions where polarity changes and where the dense wind originates \citep[e.g.][]{Phillips.etal:95,McComas.etal:07}. Changing $m$ corresponds to changing the distribution of the polarity changing rings and, thus, the latitude at which the mass will be lost.  Mass loss depends on the wind speed and density but it does not depend on the latitude where the loss happens.  Of course, if the loss would happen at higher density regions, then there would be a difference. But the density at the base of a streamer actually depends on the wind speed itself.  Areas of closed field lines (dead zones) are typically very dense, while areas of open field lines are less dense with a wind that is faster the further from the dead zone.  Changing $m$ at constant $n$ is equivalent to rearranging those rings of wind, but the total area of open field lines and the wind speed-density relation remains constant. In summary, what matters is the flux emerging from the areas of open field lines, i. e. the open flux, as discussed by \cite{Reville.etal:15a} and  \cite{Garraffo.etal:15} . 

In contrast, AML will certainly depend on the latitude of the streamers and, therefore, some variability in angular momentum loss is expected from the change in the latitude of mass loss rates for different $m$. It is worth noticing that even if the number of regions of open flux increases with $n$, the total open flux and the open flux area both decrease with complexity. 

For a given magnetic moment, the mass loss and angular momentum loss rates are independent of the choice of $m$, which means they are independent of the way in which magnetic flux is distributed for a given magnetic complexity $n$.  In the bottom panel of Figure~\ref{fig:aml} we show that this is the case for the open flux as well with the same small variability that angular momentum loss shows.

\subsection{Morphology Term} 
\label{sec:morphterm}

Our simulations, together with the ones in CG15 ($73$ simulations) , allows us to gain some insights into how the relations between magnetic flux, dipolar field strength, and mass and angular momentum loss rates get affected by magnetic morphology.  We have derived analytical approximations for the model mass and angular momentum loss results in terms of scaling factors, $Q_{M}(n)$ and $Q_{J}(n)$, that can be applied to the respective dipolar field loss rates to convert them to rates appropriate to higher order fields with equivalent total magnetic flux.   These were performed using solar parameters (base density, rotation, stellar radius, and stellar mass) and, therefore, should be valid in that regime. We expect them, however, to be valid in a wider rotation rate ($P$ larger than $\sim 5$ days), with the angular momentum loss scaling with $\Omega$.  Beyond that limit, in the fast rotation regime, there are qualitative differences in the solutions, like wrapping of the field lines, that can result in different modifications in the mass angular momentum loss rates, as discussed by \cite{Cohen.Drake:14}. In addition, beyond a certain rotation rate, magneto-centrifugal effects play an important role and should be taken into account (see also \citealt{Matt.etal:12} and \citealt{Reville.etal:15a}). 
The relations are shown in Figure~\ref{fig:aml} together with the simulation results and read
\begin{flalign}
&\dot{M}=\dot{M}_{Dip}  Q_{M}(n)&
\end{flalign}
\begin{flalign}
&Q_{M}(n)= (20/B)^{(n-1)/20}e^{(1.22-1.42 n + 0.19n^2 + 0.01n^3)}&
\end{flalign}
\begin{flalign}
&\dot{J} =  \dot{J}_{Dip} Q_{J}(n) & 
\end{flalign}
\begin{flalign}
&Q_{J}(n) = 4.05 \, e^{-1.4 n}+(n-1)/(60 B\, n) &
\end{flalign}\\
Here, $\dot{M}_{Dipole}$ and $ \dot{J}_{Dipole}$ correspond to the mass and angular momentum loss rates assuming a dipolar magnetic morphology, $Q_{M}(n)$ and $Q_{J}(n)$ are the terms representing the magnetic morphology dependence, $B$ stands for field strength [Gauss], and $n$ stands for the magnetic multipolar moment (level of complexity of the field).

\subsection{Real Stars}
\label{sec:realstars}

So far we have only considered pure modes of the spherical harmonics decomposition. It would be useful to understand how this analysis applies to real stars. Any magnetogram is a linear combination of pure modes, however the response of the plasma to the magnetic field is non-linear and, therefore, the stellar corona will not be a simple superposition of the solutions for each mode.  We have discussed here that mass loss and angular momentum loss rates are modulated by the level of complexity $n$ of the field for pure modes and that the dipolar solution is not always a good approximation for complex morphologies. Therefore, keeping in mind that stellar morphology can not be fully described by a simple index, it would still be convenient to find a parameter that best represents complexity.  For real magnetograms we found such an index that allow us to analytically estimate the mass and angular momentum loss rates much more realistically than when assuming dipolar morphology. 

\begin{figure*}[h]
\center
 \includegraphics[trim = 0.3in 0.2in
  0.1in 0.01in,clip, width = 0.4 \textwidth]{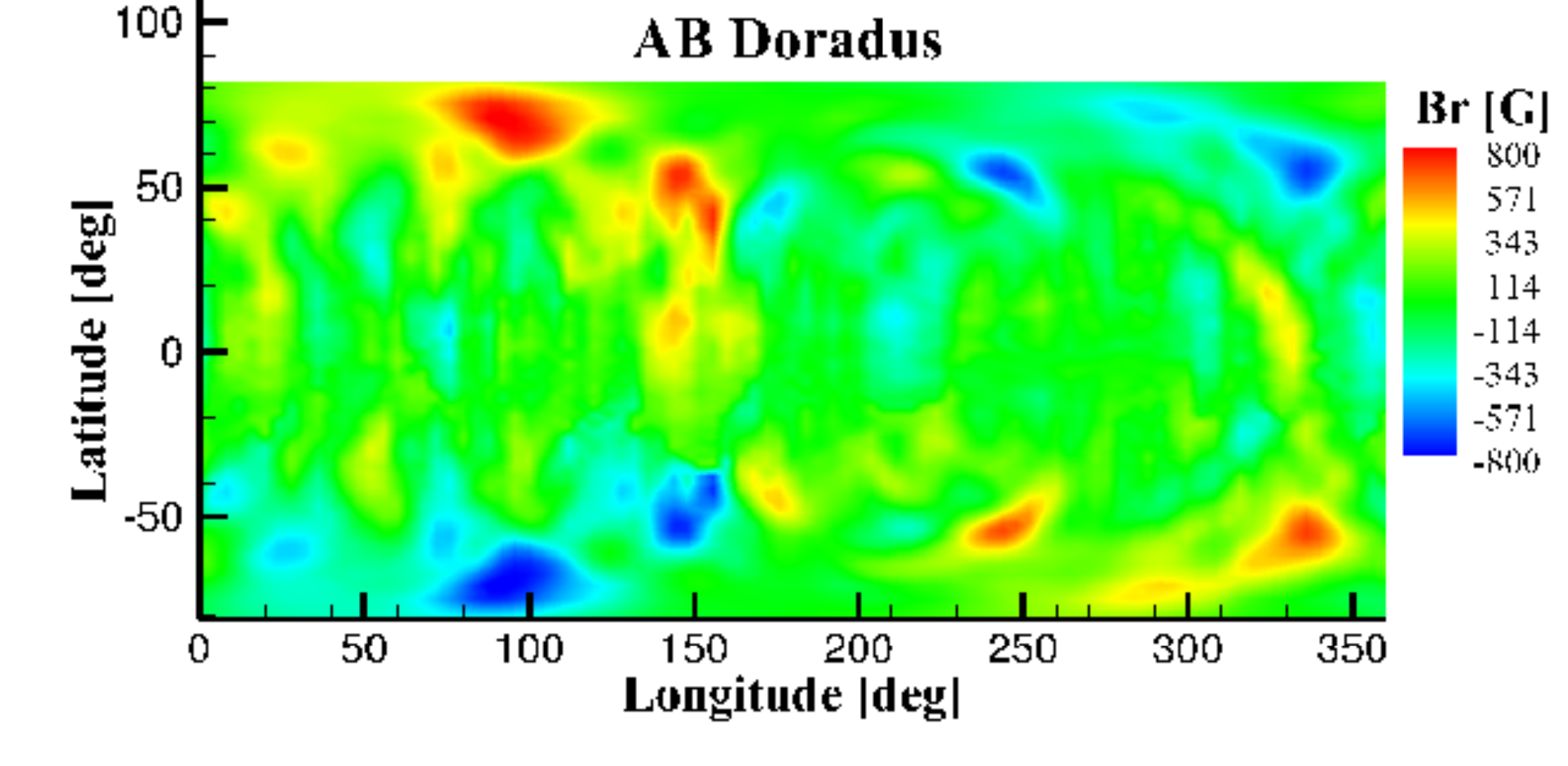} 
 \includegraphics[trim = 0.3in 0.2in
  0.1in 0.01in,clip, width = 0.4 \textwidth]{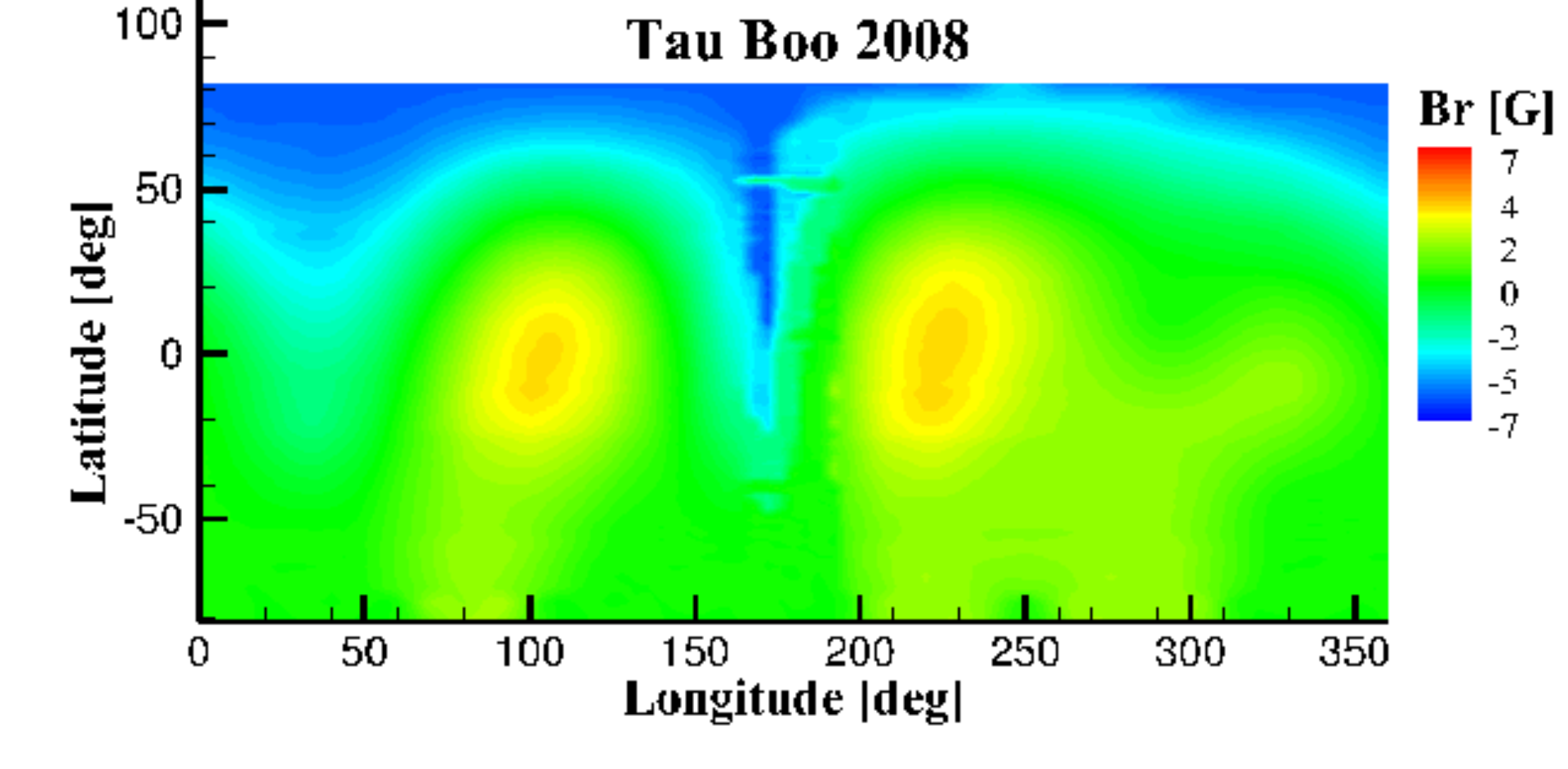} \\
 \includegraphics[trim = 0.3in 0.2in
  0.1in 0.01in,clip, width =  0.4 \textwidth]{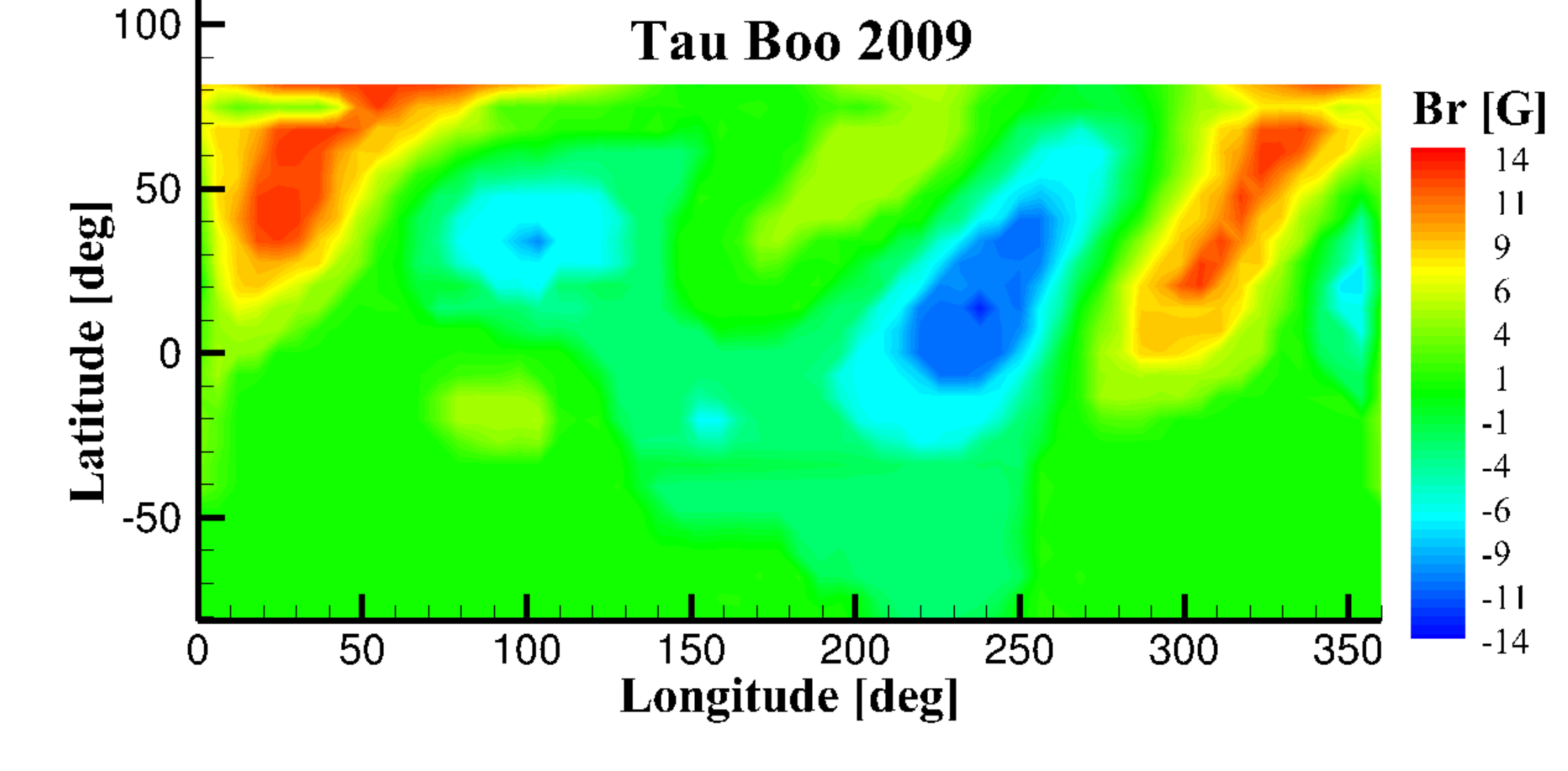} 
 \includegraphics[trim = 0.3in 0.2in
  0.1in 0.01in,clip, width =  0.4 \textwidth]{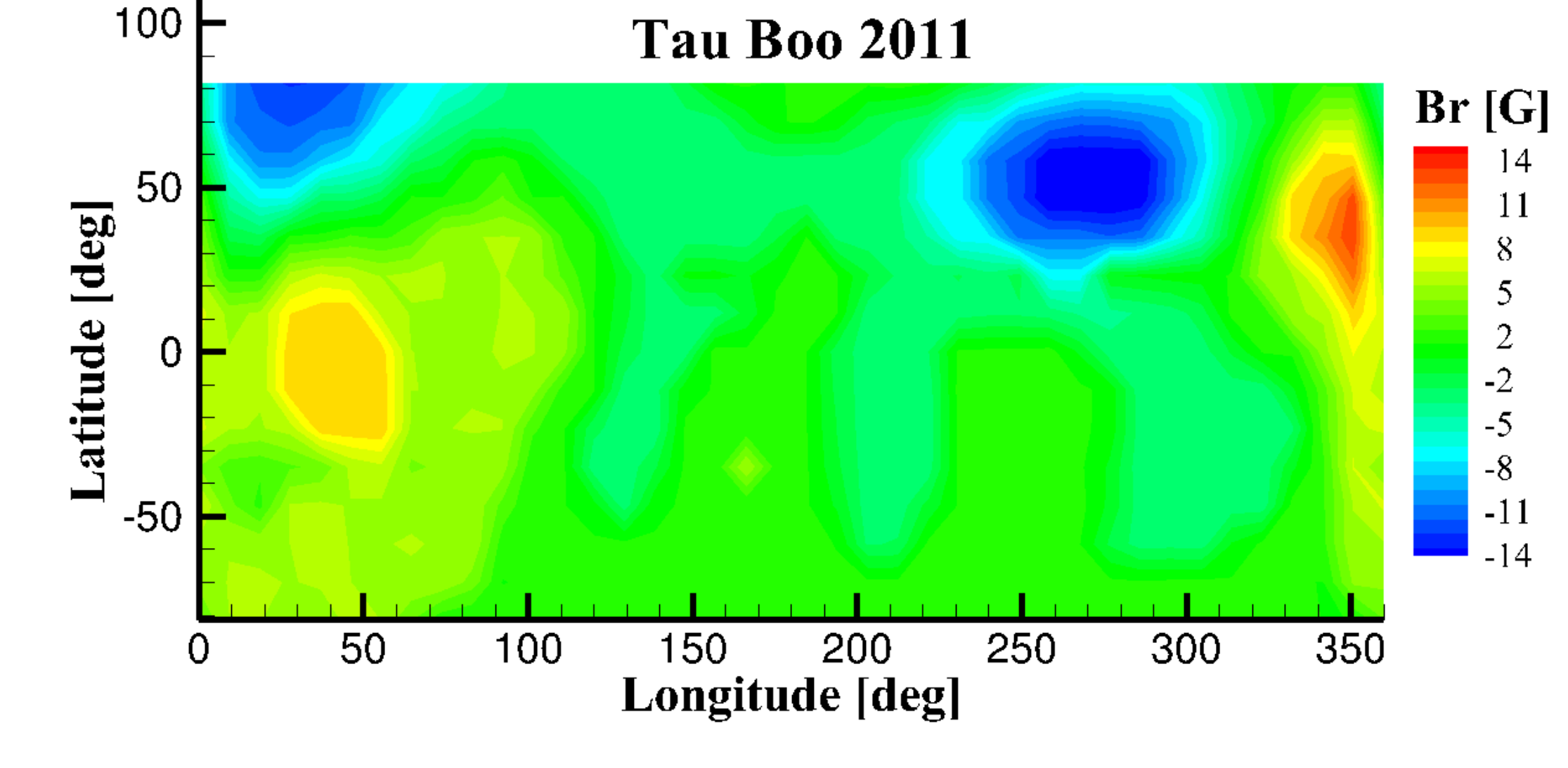} \\
\includegraphics[trim = 0.3in 0.2in
  0.1in 0.01in,clip, width =  0.4 \textwidth]{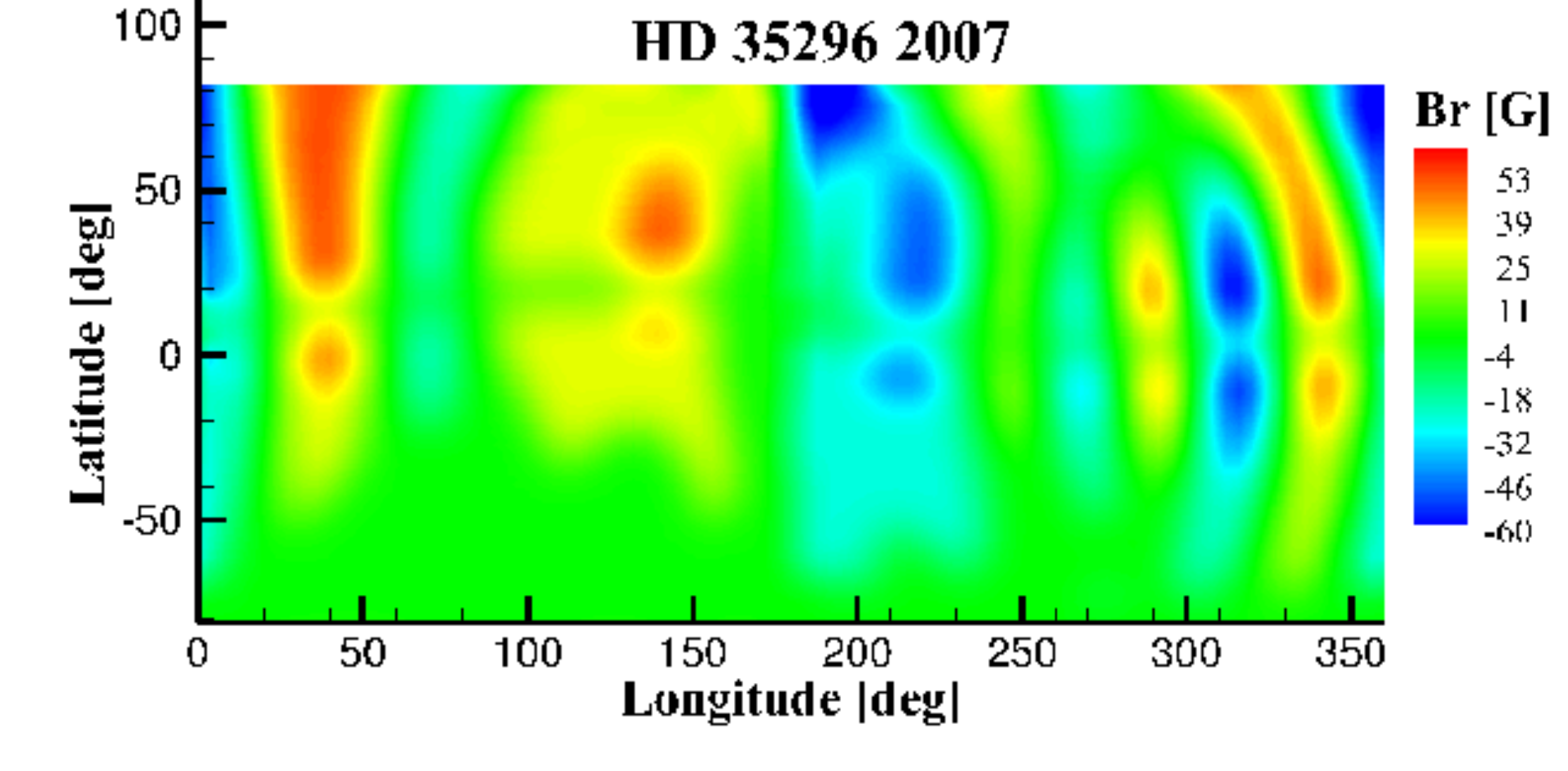} 
\includegraphics[trim = 0.3in 0.2in
  0.1in 0.01in,clip, width =  0.4 \textwidth]{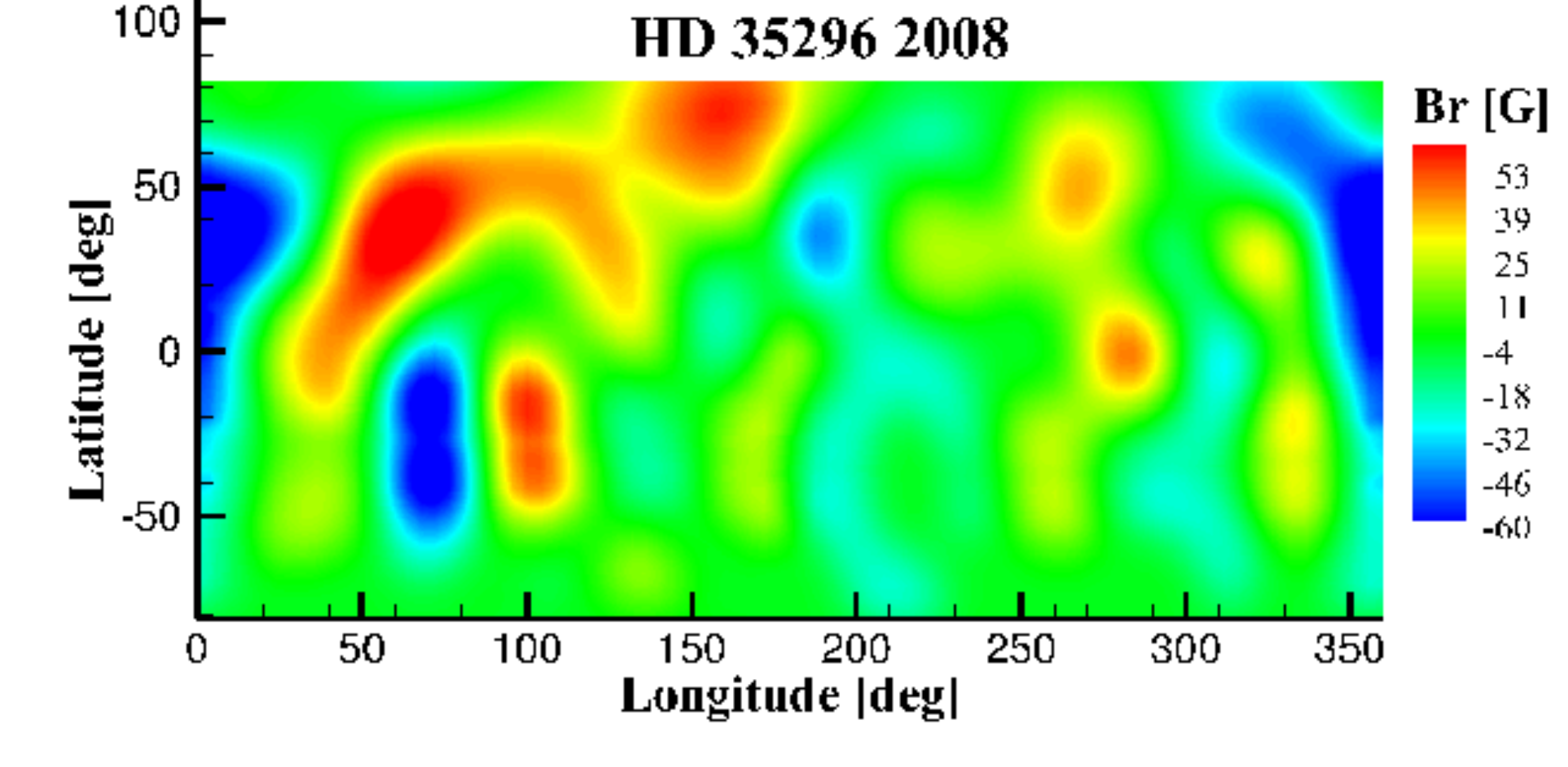} \\
 \includegraphics[trim = 0.3in 0.2in
  0.1in 0.01in,clip, width =  0.4 \textwidth]{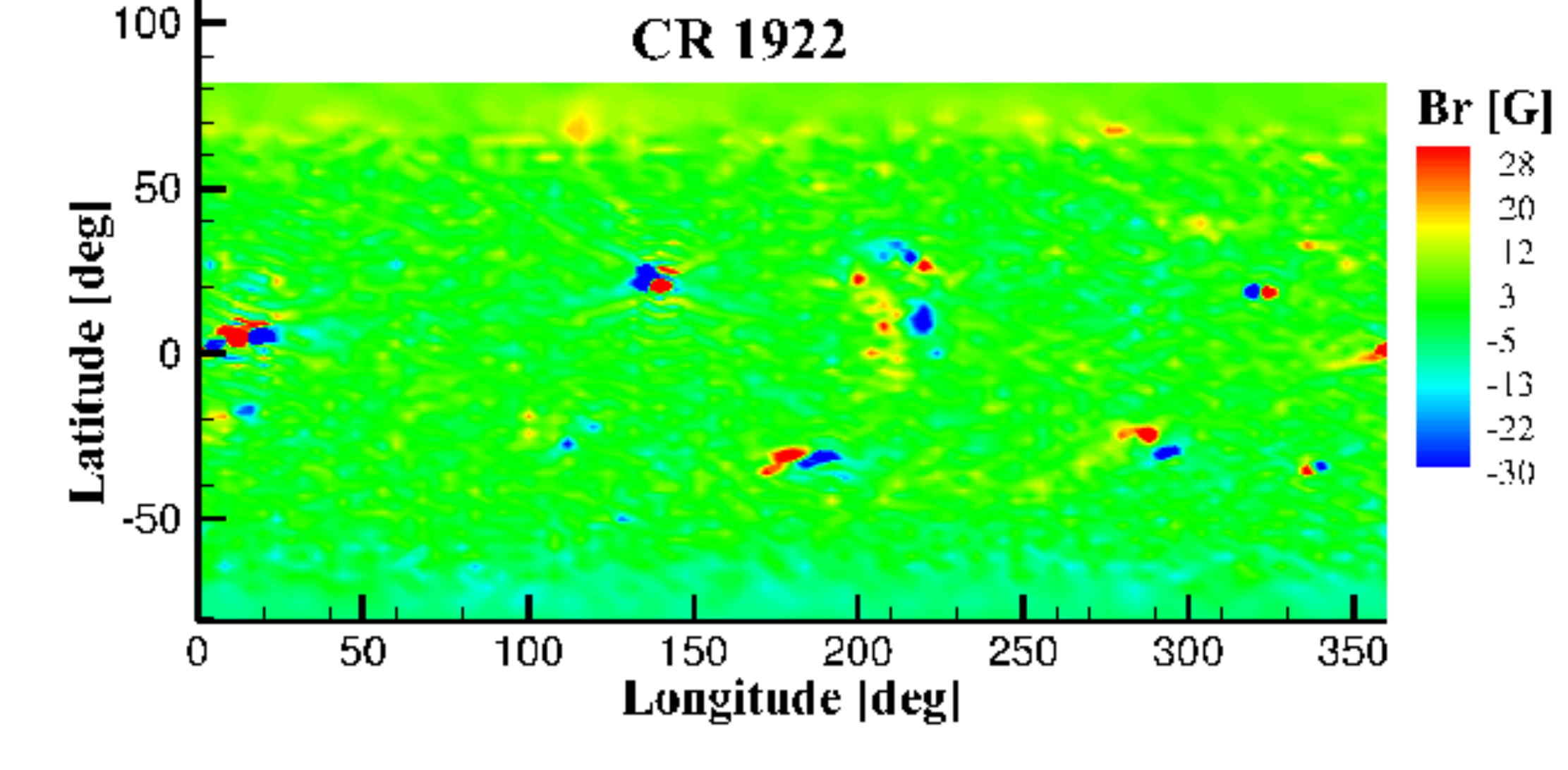} 
\includegraphics[trim = 0.3in 0.2in
  0.1in 0.01in, clip, width = 0.4 \textwidth]{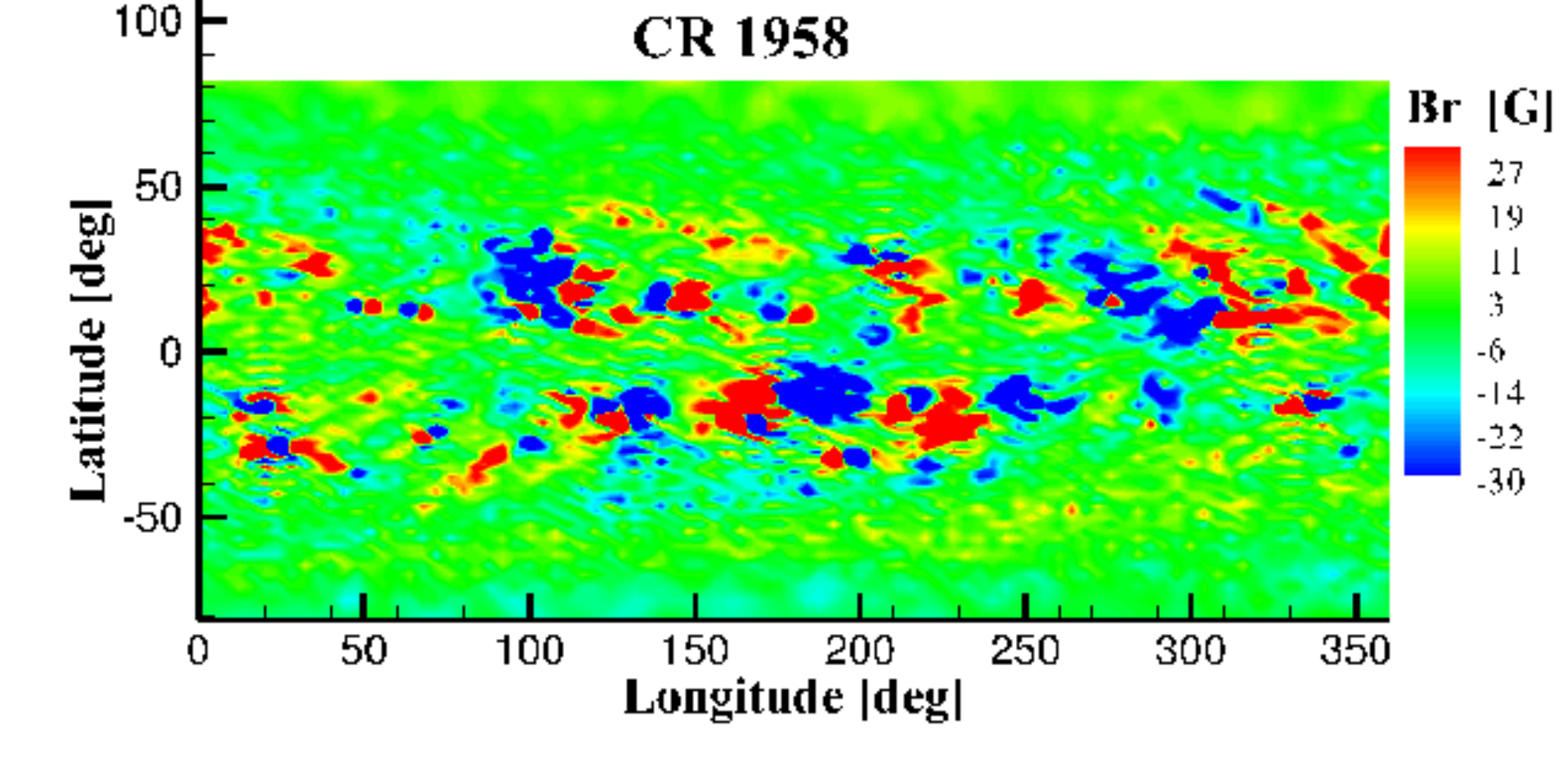} 
  \caption{ZDI observed magnetograms for our sample of real stars.}
\label{fig:realmagnetograms}
\end{figure*}

We use $8$ ZDI observations of solar-like stars (see Figure~\ref{fig:realmagnetograms}): AB Doradus \citep{Hussain.etal:07}, $\tau$ Boo (2008 observation by \citealt{Fares.etal:09}, and 2009 and 2011 observations by \citealt{Fares.etal:13}), HD 35296 (2007 and 2008 observations by \citealt{Waite.etal:15}), and solar maximum (CR 1958) and solar minimum (CR 1922) obtained by the Solar and Heliospheric Observatory (SoHO) Michelson Doppler Imager (MDI)\footnote{http://sun.stanford.edu}.  For each star in our sample we decompose the magnetogram in spherical harmonics and calculate a complexity parameter as the magnetic flux weighted average of the magnetic multipolar order:
\begin{equation}
\, \, \, n_{av} = \sum_{n=0}^{n_{max}} \frac{n \, F_n}{F_T},
\label{eq:nav}
\end{equation}
where $F_n$ is the magnetic flux in each term in the decomposition and $F_T$ is the flux in the original magnetogram. 

Using our analytical expressions we obtain, from the total magnetic flux and the complexity parameter $n_{av}$, the expected mass and angular momentum loss rates for each star. We performed simulations for the $8$ magnetograms in our sample, using the solar parameters described above (base density, rotation period, stellar radius, and stellar mass) in order to be consistent with our theoretical analysis, and compared the mass and angular momentum loss rates with the analytical predictions.  As we mentioned above, coronae solutions cannot be linearly added and, therefore, differences between the simulated solutions and the analytical estimates are expected.  The idea here is to find a proxy that provides a good enough estimate of complexity and that will represent an improvement over the usual dipolar assumption in spin-down models. We do not claim these estimates to be exact solutions. It is for that reason that we test our estimates in the $8$ real cases shown in Figure~\ref{fig:realmagnetograms}.

\begin{figure*}[]
\resizebox{\hsize}{!}{\includegraphics[trim = .8in 5.in 
.9in 1.5in,clip,width = \textwidth]{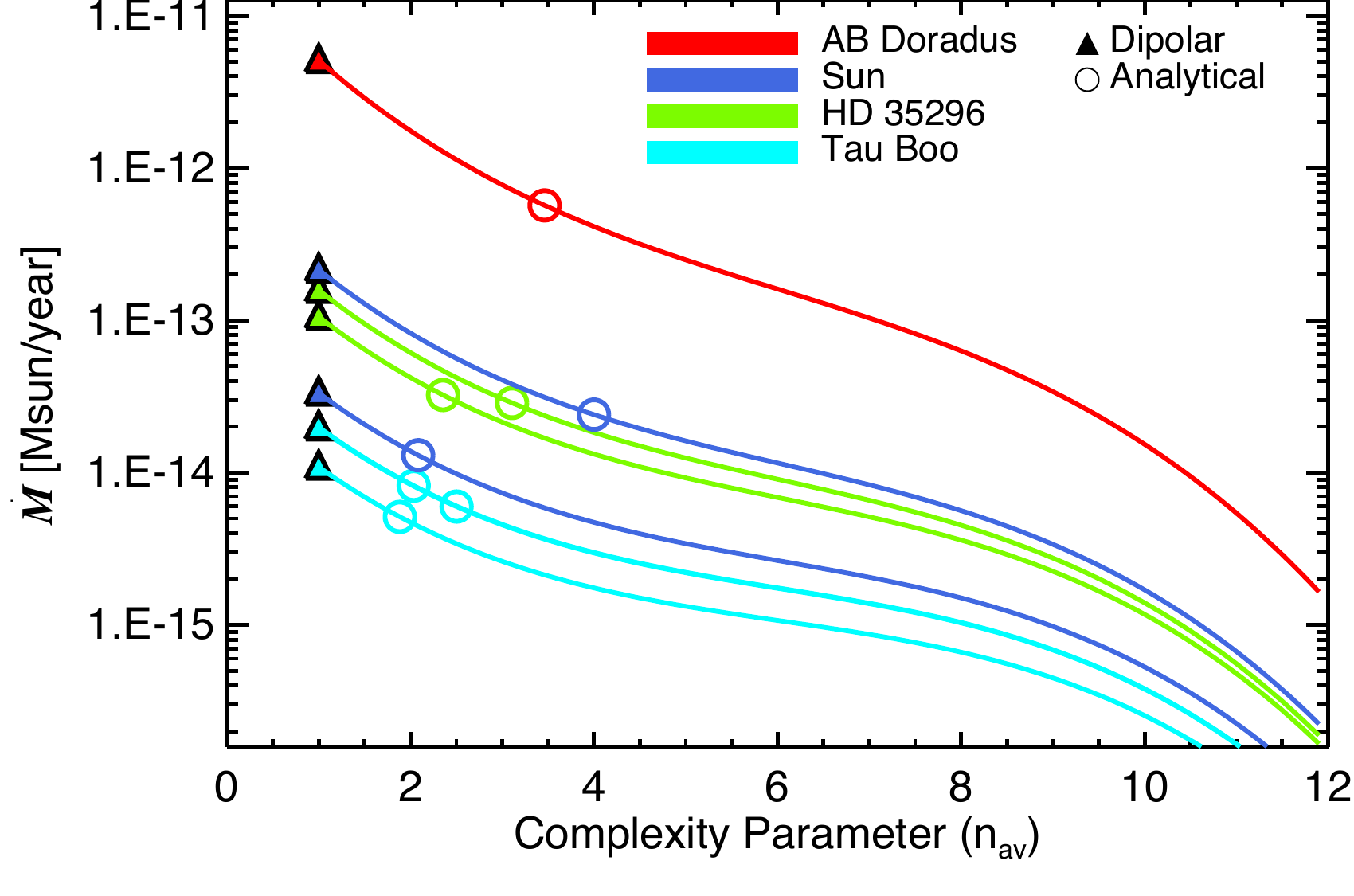}
\includegraphics[trim = .8in 5.in 
.9in 1.5in,clip,width = \textwidth]{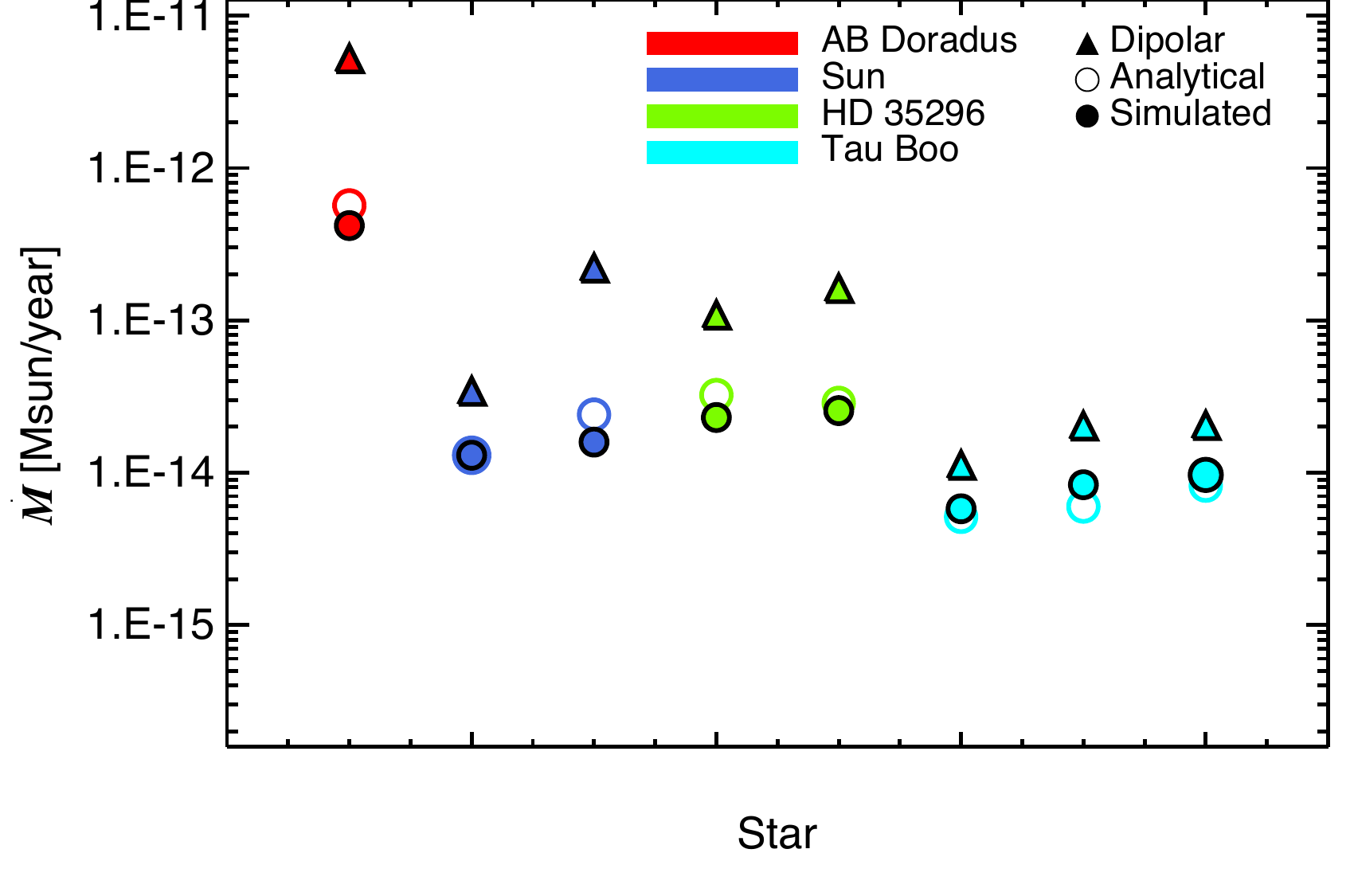} }
\resizebox{\hsize}{!}{\includegraphics[trim = .8in 5.in 
.9in 1.5in,clip,width = \textwidth]{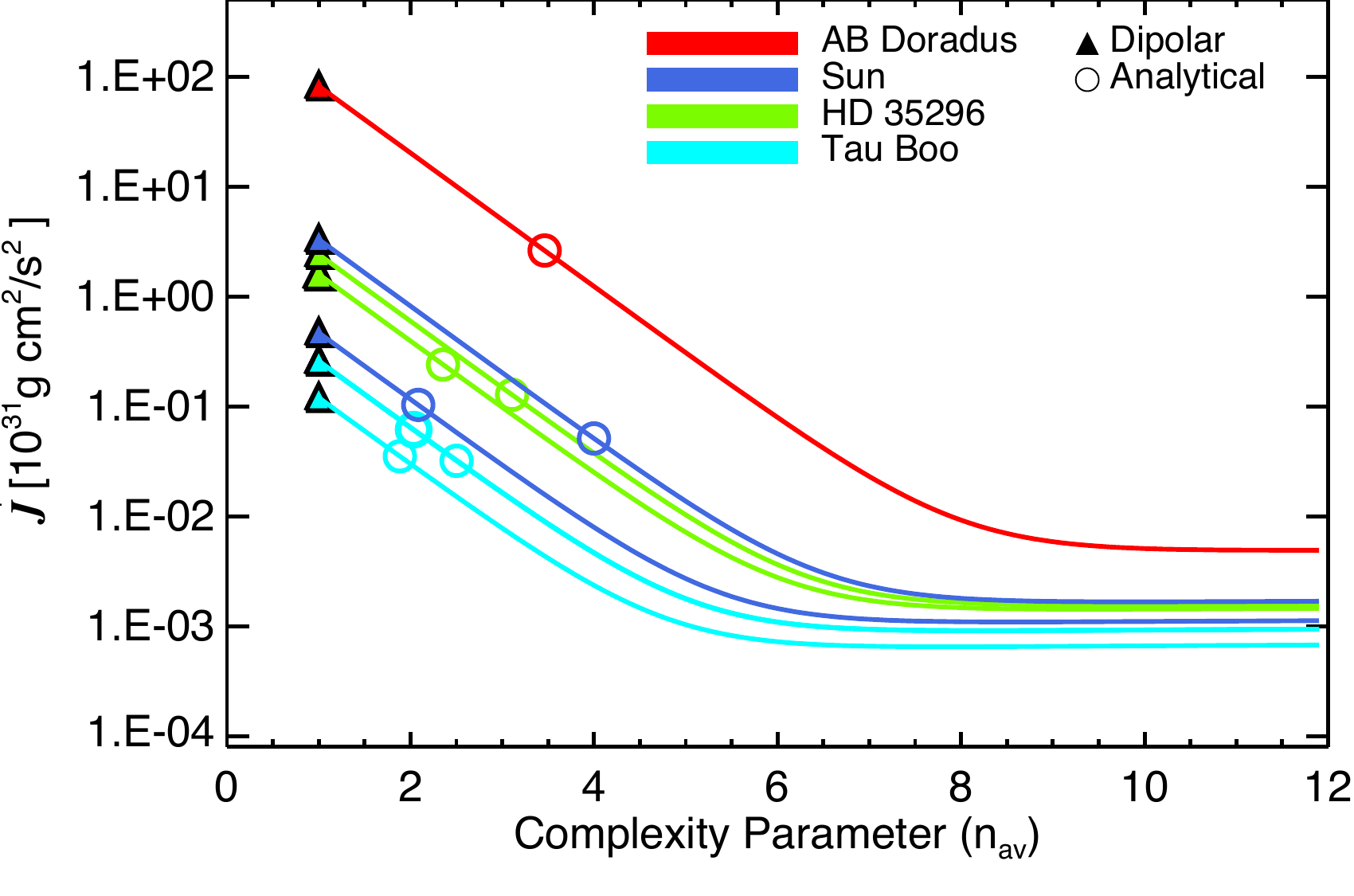}
\includegraphics[trim = .8in 5.in 
.9in 1.5in,clip,width = \textwidth]{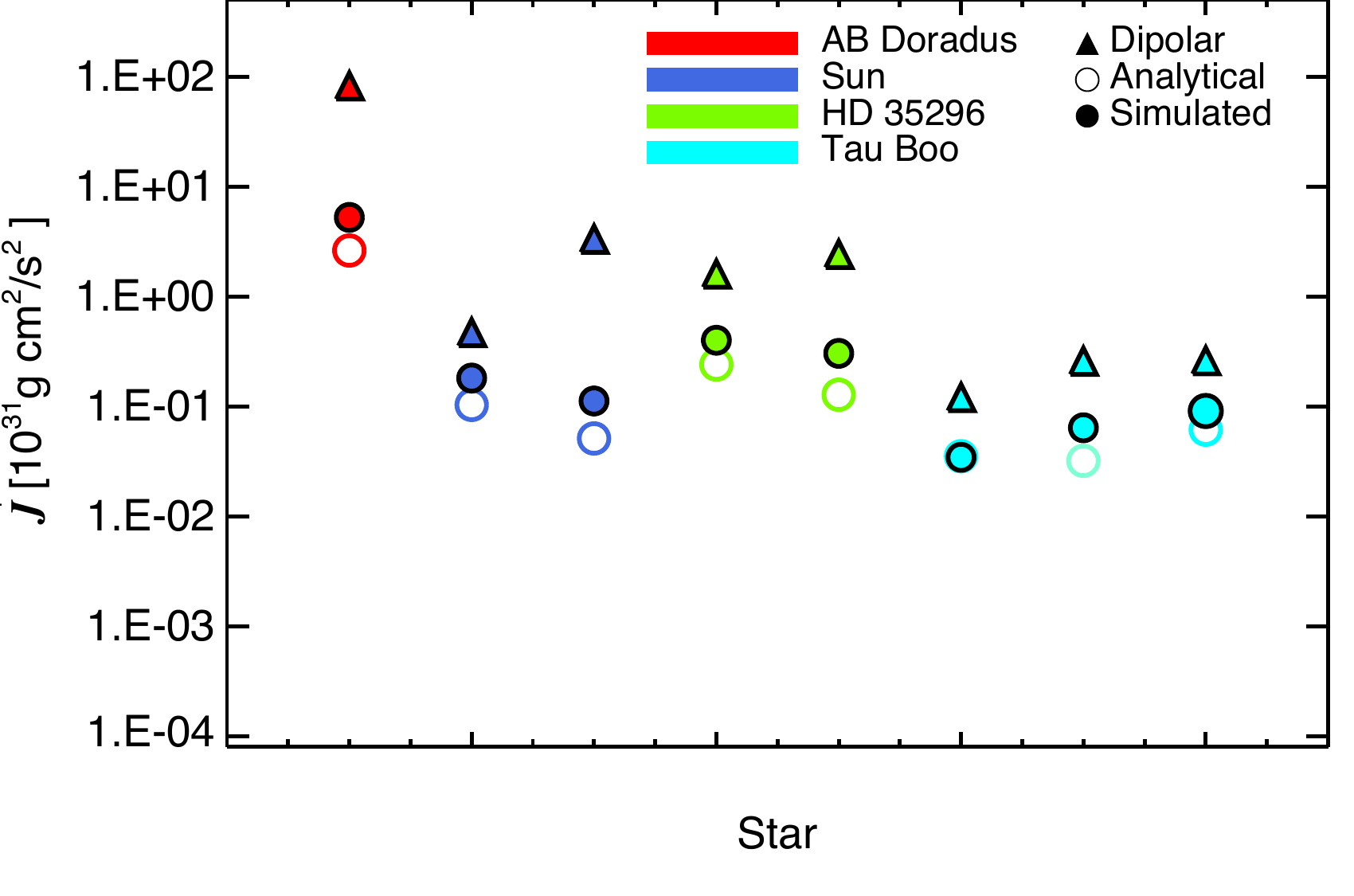} }
\caption{ Mass ($\dot M$ top) and angular momentum ($\dot J$ bottom) loss
  rates for our sample of $8$ observed magnetograms (see Figure~\ref{fig:realmagnetograms}).  The left panels show the analytical estimations as described in Section~\ref{sec:realstars}. Triangles show dipolar morphology of the same total magnetic flux , the scaling laws are shown as lines,  and the resulting analytical estimations (for the corresponding $n_{av}$) are represented by hollow circles. The right panels show the comparison between the equivalent dipolar rates (represented by triangles), the estimated rates (represented by hollow circles), and the results from our simulations (represented by full circles).} 
\label{fig:relations}
\end{figure*}

Figure~\ref{fig:relations} shows our results, the top panel is a comparison of the simulated and analytically predicted mass loss rates and the bottom panel the comparison of the simulated and analytically predicted angular momentum loss rates.  One can see that the estimations are consistently similar to the simulated results and, therefore, the complexity parameter defined as above is a representative one 
that can be used to estimate mass loss and spin down rates solely from the total flux and morphology of the magnetic field.  

For the sun, our predictions as well as the results from our simulations are consistent with the inferred solar mass loss
rate of $2 \cdot 10^{14} M_{\sun} yr^{-1}$, derived from typical solar wind
parameters near Earth, and with previous studies showing that the mass loss rate of the sun is constant through its magnetic cycle (see \citealt{Cohen:11}  and references therein).  It is worth noticing in the case of solar maximum that, while $n \sim 4$ gives a much smaller angular momentum rate than a dipole it is compensated by the fact that the magnetic flux is larger by a factor of $\sim 3$ than the flux in solar minimum.
 We conclude that these represent a much more realistic scenario than the simplistic assumption of dipolar morphology, and that they provide an important new ingredient to rotation evolution models.
  
\vspace{2in}  
   
\section{CONCLUSIONS}
\label{sec:Conclusions}

Our MHD simulations indicate that mass and angular momentum loss rates from solar-like stars are strongly suppressed by magnetic complexity but independently of how this complexity is organized.  The loss rates are controlled by the number of rings in which the polarity changes sign, labeled by the spherical harmonic order of complexity, $n$.  Translating these results to non-ideal surface field distributions of real stars, both mass and angular momentum loss rates depend on the level of complexity of the field only and not on the details of the field distribution over the stellar surface. 
We have provided analytical formulae for the dependence of mass and angular momentum loss rates on magnetic complexity that can be applied in stellar rotation evolution models.  We have shown that this ingredient significantly improves the analytical estimations of mass and angular momentum loss rates based on ZDI maps (total magnetic flux and complexity) over the usual dipolar assumption. 



\begin{acknowledgements}
CG and OC were supported by SI Grand Challenges grant ``Lessons from Mars: Are Habitable Atmospheres on Planets around M Dwarfs Viable?''.   OC was also supported by SI CGPS grant ``Can Exoplanets Around Red Dwarfs Maintain Habitable Atmospheres?'' and "Living With a Star" grant NNX16AC11G.  JJD was supported by NASA contract NAS8-03060 to the {\it Chandra X-ray Center}. The authors thank Vinay Kashyap for helpful discussions.  
Numerical simulations were performed on the NASA HEC Pleiades system under award SMD-13-4526, and on the Smithsonian Institution High Performance Cluster (SI/HPC).
\end{acknowledgements}


\end{document}